\documentclass{emulateapj}
\newcommand\etal{{\it et al.~}}
\newcommand\beq{\begin{equation}}
\newcommand\eeq{\end{equation}}
\newcommand\beqar{\begin{eqnarray}}
\newcommand\eeqar{\end{eqnarray}}

\newcommand\bvec[1]{\hbox{\boldmath${#1}$}}

\newcommand{\vef}{v_{\rm e\phi}}
\newcommand{\vif}{v_{\rm i\phi}}
\newcommand{\vgof}{v_{\rm g_0\phi}}
\newcommand{\vgpf}{v_{\rm g_+\phi}}
\newcommand{\vgmf}{v_{\rm g_-\phi}}
\newcommand{\vi}{v_{\rm i}}
\newcommand{\ve}{v_{\rm e}}
\newcommand{\vn}{v_{\rm n}}
\newcommand{\vgo}{v_{\rm g_0}}
\newcommand{\vgp}{v_{\rm g_+}}
\newcommand{\vgm}{v_{\rm g_-}}
\newcommand{\vf}{v_{\rm f}}

\newcommand{\tz}{\tau_0}
\newcommand{\tp}{\tau_{\rm +}}
\newcommand{\tm}{\tau_{\rm -}}

\newcommand{\tgn}{\tau_{\rm gn}}
\newcommand{\rgo}{\rho_{\rm g0}}
\newcommand{\rgp}{\rho_{\rm g+}}
\newcommand{\rgm}{\rho_{\rm g-}}
\newcommand{\oitin}{\omega_{\rm i} \tau_{\rm in}}
\newcommand{\ogtgn}{\omega_{\rm g}\tau_{\rm gn}}
\newcommand{\oeten}{\omega_{\rm e} \tau_{\rm en}}

\newcommand{\rmm}{\frac{\rgo}{\rgm}\frac{\tgn}{\tm}}
\newcommand{\rpp}{\frac{\rgo}{\rgp}\frac{\tgn}{\tp}}

\newcommand{\tref}{T_{\rm iso}}

\newcommand{\trot}{T_{\rm rot}}

\newcommand{\tildpext}{\tilde{P}_{\rm ext}}
\newcommand{\cref}{C_{\rm ref}}

\newcommand\tens[1]{\overline{\overline{#1}}}

\begin{document}

\title{PROTOSTAR FORMATION IN MAGNETIC MOLECULAR CLOUDS BEYOND ION DETACHMENT:
\\I. FORMULATION OF THE PROBLEM AND METHOD OF SOLUTION} 

\author{Konstantinos Tassis\altaffilmark{1,2} \& Telemachos Ch. Mouschovias\altaffilmark{1}}

\altaffiltext{1}{Departments of Physics and Astronomy,
University of Illinois at Urbana-Champaign, 1002 W. Green Street, Urbana, IL 61801}
\altaffiltext{2}{Department of Astronomy and Astrophysics and the 
Kavli Institute for Cosmological Physics, 
The University of Chicago, 5640 South Ellis Avenue,
Chicago, IL 60637}

\begin{abstract}

We formulate the problem of the formation of magnetically supercritical cores in magnetically 
subcritical parent molecular clouds, and the subsequent collapse of the cores to high densities, 
past the detachment of ions from magnetic field lines and into the opaque regime. We employ the 
six-fluid MHD equations, accounting for the effects of grains (negative, positive and neutral) 
including their inelastic collisions with other species. We do not assume that the magnetic flux 
is frozen in any of the charged species. We derive a generalized Ohm's law that explicitly 
distinguishes between flux advection (and the associated process of ambipolar diffusion) and 
Ohmic dissipation, in order to assess the contribution of each mechanism to the increase of 
the mass-to-flux ratio of the central parts of a collapsing core and possibly to the resolution 
of the magnetic flux problem of star formation. 
The results, including a detailed parameter study, are presented in two accompanying papers. 
\end{abstract}

\keywords{ISM: clouds -- ISM: dust -- magnetic fields -- MHD -- stars: formation -- shock waves}

\section{Introduction: Observational and Theoretical Background}

In the Milky Way, stars are observed to form inside molecular clouds, 
located primarily along spiral arms. Molecular clouds do not exhibit 
significant ordered expansion or contraction velocities. They are 
however observed to have supersonic but subAlfv\'{e}nic linewidths. 
Polarization \citep{Hildeb99,Rao98,Lai02} and Zeeman \citep{Crut87,
Crut94} observations have revealed that such clouds are threaded by 
ordered magnetic fields, of magnitude sufficient to support them 
against their self-gravity.

In such magnetically supported ({\em magnetically subcritical}) clouds, 
magnetically supercritical fragments (or cores) form by the process 
of ambipolar diffusion, which allows neutral molecules to diffuse 
through charged particles and magnetic field lines, on a timescale 
1 - 10 Myr (depending on the local physical conditions). When self-gravity 
becomes strong enough to overwhelm the magnetic forces, the fragments 
begin to contract dynamically. The initial stages of this contraction 
are isothermal. However, the column density eventually increases enough 
that radiation is trapped, the temperature rises, and hydrostatic 
protostellar cores form in the fragments. These late stages of contraction 
and the formation of hydrostatic protostellar cores are the subject 
matter of this paper. 

The leftover magnetic flux at the end of the early, isothermal phase of
contraction ($n \lesssim 10^{10}$ ${\rm cm^{-3}}$) exceeds observed
stellar magnetic fluxes by two to four orders of magnitude \citep{m87b} 
and must therefore be redistributed or dissipated during the 
pre-main--sequence evolution. Approximate theoretical estimates concluded that
Ohmic dissipation is effective in destroying the magnetic flux at
densities above $10^{10}$ ${\rm cm^{-3}}$ \citep{pm65,nnu91}.
\cite{pm65} considered both ambipolar diffusion and Ohmic
dissipation as magnetic flux-loss mechanisms and performed detailed
calculations of the charged species abundances, accounting for ionization
due to cosmic rays, radioactivity and thermal ionization. They
estimated the relative importance of the flux-reduction mechanisms by
comparing their timescales to the free-fall timescale, and 
concluded that Ohmic dissipation is more effective than ambipolar diffusion
at densities $\gtrsim 10^{10} \, {\rm cm ^{-3}}$. The field is 
recoupled to the matter when, at high enough temperatures, thermal 
ionization raises the degree of ionization sufficiently. They estimated the
magnetic field of a newly born solar-mass star to be $\sim 40$ ${\rm
Gauss}$.  

Two decades later \cite{nu86a, nu86b} repeated these calculations using 
a more refined chemical model. Through timescale comparisons for a
series of different geometries (disk-like or
spherical), they also found that the magnetic field decouples from the 
matter because of Ohmic dissipation. This decoupling took place at densities
between $10^{11}$ and $10^{12} {\, \rm cm^{-3}}$. Further refinements
in the chemistry \citep{UN90} and the grain size distribution \citep{nnu91}
yielded similar results.

Current-driven instabilities may cause anomalous magnetic-flux destruction
\citep{NH85} above densities $10^{11} {\, \rm cm^{-3}}$. Such instabilities, 
however, require drift velocities of charged species greater 
than their thermal velocities, a condition not realized 
in any of the dynamical calculations \citep{FM92,FM93,CM93,CM94,CM95,BM94}. 

Recent dynamical calculations by \cite{DM01}, 
which follow the evolution up to densities $n = 2 \times 10^{12}$ ${\rm
  cm^{-3}}$, show that ambipolar diffusion is more effective than Ohmic
dissipation in decoupling the magnetic field from the neutral
matter. [Earlier calculations \citep{nu86a, nu86b, UN90, nnu91} overestimated 
the ${\rm e-H_2}$ elastic
collision cross section, and hence the Ohmic resistivity, by a factor
of 100; see review by \cite{m96}].

In this series of papers, we pursue the self-initiated, 
ambipolar-diffusion--controlled star formation to densities 
$\approx 10^{15} {\, \rm cm^{-3}}$, by which opaque hydrostatic 
protostellar cores form. It is in this density regime (after the 
detachment of ions and before thermal ionization becomes important 
and the increased degree of ionization reattaches the matter to the 
field) that the field may become completely detached from the matter. 
A significant contribution to the resolution of the magnetic flux 
problem of star formation is expected at this stage. Whether ambipolar 
diffusion or Ohmic dissipation is more effective in increasing the 
mass-to-flux ratio can only be determined through such detailed evolutionary 
calculations, not through comparisons of timescales.

In the present paper we formulate the problem in terms of the six-fluid MHD equations 
(neutral molecules; atomic and molecular ions; electrons; negative, positive and neutral grains). 
We do not assume that the magnetic flux is frozen in any fluid (in order to account for the 
possibility that all charges become detached from the magnetic field). We derive a new expression 
for the generalized Ohm's law, which lies at the heart of our approach to the problem at hand.
This new formulation (1) includes the effects of inelastic coupling between 
grain species, and (2) explicitly distinguishes between flux advection, ambipolar diffusion, 
and Ohmic dissipation. We show how our formulation is related to and can be transformed into the 
traditional, ``directional'' formulation of the generalized Ohm's law, and we derive formulae
for the perpendicular, parallel and Hall conductivities entering the latter, which include, for the 
first time, the effect of inelastic collisions between grains. In addition, we present a general 
(valid in any geometry) solution for the velocities of charged species as functions of the 
velocity of the neutrals and of the effective flux velocity (which can in turn be calculated from the 
dynamics of the system and Faraday's law). The last two sets of formulae can be adapted for 
use in any general MHD code to study phenomena beyond star formation in magnetic clouds.

Our results for a fiducial model cloud and a detailed parameter study are presented in two 
companion papers (Tassis \& Mouschovias 2006b, c; hereafter papers II and III, respectively).
As shown in paper III, at the high densities treated here
memory of the mass-to-flux ratio of the parent cloud is lost, and the 
evolution of the contracting core is essentially  
independent of the initial conditions (for values within 
the parameter space determined by observations). This point is of particular 
importance, especially in light of the recent debate concerning the formation 
processes of molecular cloud cores. 
(For reviews and detailed discussion of the arguments on both sides of this debate, see, for example, Elmegreen \& Scalo 2004; 
MacLow \& Klessen 2004; Tassis \& Mouschovias 2004; Scalo \& Elmegreen 2004; Ballesteros-Paredes \& Hartmann 2006; and Mouschovias, Tassis \& Kunz 2006 and references therein).  
The need for a nonideal MHD treatment of the high-density 
regime of core evolution, when the magnetic flux problem of star formation has to 
be resolved, and hence the relevance of this work, {\em is 
independent of the adopted theory of core formation.}

\section{Model Cloud}

We consider a self-gravitating, magnetic, weakly-ionized model molecular cloud
consisting of neutral particles (${\rm H_2}$ with $20\%$ ${\rm He}$
by number), ions (both molecular ${\rm HCO^+}$ and atomic ${\rm
  Na^+}$, ${\rm Mg^+}$), electrons, neutral, singly negatively-charged
and positively-charged grains.
The abundances of charged species are determined from the chemical reaction
network discussed in Tassis \& Mouschovias (2005a, Appendix A). For a detailed 
discussion of the chemistry,
see \cite{CM93,CM95}, and \cite{DM01}. 
Cosmic rays of energy $\gtrsim 100 {\, \rm MeV}$, able to penetrate
deep inside the cloud, drive the ionization for a wide range of densities.
However, cosmic rays are 
appreciably attenuated for column densities $\gtrsim 100 {\rm \, g \,
  cm^{-2}}$. At even higher densities, radioactive decays become the dominant source of
ionization \citep{pm65,UN81, U83}. The ionization rate $\zeta$ is the sum of the 
cosmic-ray ionization rate ($\zeta_{\rm CR}$) and the rate 
of ionization due to natural radioactivity ($\zeta_{\rm RD}$). 
The cosmic-ray ionization rate can be described \citep{UN80} by the relation: 
\beq
\zeta_{\rm CR} = \zeta_{\rm CR,0} \exp({-\sigma / 96 \,\, {\rm g \, cm^{-2}}}),
\eeq
where $\sigma$ is the column density of the gas, and $\zeta_{\rm CR,0}$ is the unattenuated 
cosmic-ray ionization rate.
Values of $\zeta_{\rm CR,0}$ are estimated to lie in the range $10^{-17}$ - $10^{-15}$ ${\rm s^{-1}}$,
with  $5 \times 10^{-17}$ ${\rm s^{-1}}$ accepted as a canonical value \citep{sp78}.
The ionization rate due to natural radioactivity is dominated by the the isotope $ ^{40}{\rm K}$ 
because of its long half-life of $1.25$ Gyr. In the present calculations $\zeta_{\rm RD}=6.9 \times 
10^{-23}$ ${\rm s^{-1}}$ \citep{UN80}. When the isotope $^{26}{\rm Al}$ is considered, $\zeta_{\rm RD}
=1.94 \times 10^{-19}$ ${\rm s^{-1}}$ \citep{CJ78}.
UV radiation provides an additional ionization mechanism, but it only
affects the envelope of molecular clouds \citep{CM95}. In this
paper, UV radiation is not taken into consideration, as we concentrate our
attention to the very dense, central parts of molecular clouds.

For simplicity, we consider spherical grains of uniform size, since the chemical model 
used by \cite{DM01} accounted for grains of
different sizes and showed that the effect of the grain size distribution on
the evolution of the cloud is minimal. In the
case of ion neutralization through collisions of ions 
(molecular or atomic) with grains, we assume that
the ions do not get attached to the grains, but that they get
neutralized and the resulting neutral particle escapes into the gas
phase. Thus the total abundance of metals as well as the ${\rm
  HCO}$ abundance remain constant. Grain growth and depletion of elements by
attachment onto grains are not considered here.

The axisymmetric model cloud is initially
in an exact equilibrium state with self-gravity and external pressure
being balanced by internal magnetic and thermal-pressure forces. 
We follow the ambipolar-diffusion--initiated evolution from mean molecular cloud densities 
to a stage beyond the formation of a hydrostatic protostellar core.
Isothermality is an excellent approximation for the early stages of star formation 
while the density is below $10^{10}$ ${\rm cm^{-3}}$ \citep{Gaustad63,Hayas66}. The
density at which isothermality breaks down depends on grain opacities. With a value of
$0.039$ ${\rm cm^2 \, g^{-1}}$ for spherical grains \citep{DL84}, and $0.056$ ${\rm cm^2 \, g^{-1}}$ 
for fluffy grains \citep{OH94}, the required dust column density for the trapping of radiation is 
reached at densities $\geq 10^{11}$ ${\rm cm^{-3}}$. However, the optical depth is not the determining 
factor for the breakdown of isothermality \citep{MI99}. Instead, the rate of compressional heating 
must be compared to the rate of escape of thermal radiation. With these considerations, \cite{DM01} 
found that deviations from $T=10$ K should occur at densities
$\geq 3 \times 10^{10}$ ${\rm cm^{-3}}$ but that the temperature is not expected to rise rapidly, 
since a small temperature gradient suffices to allow the energy generated by the collapse to escape.
To avoid the complicated and computationally taxing radiative transfer calculation, we switch from 
an isothermal to an adiabatic equation of state at a critical density which we treat as 
a free parameter.

\section{The Six-Fluid MHD Description of Magnetic Star Formation}

We extend the multifluid formalism described by \cite{TassisM05a} 
to relax the assumption of 
flux-freezing in the electrons and to account for the breakdown of isothermality at high densities. 
The equations governing the
behavior of the six-fluid system (neutrals, electrons, ions,
negative, positive, and neutral grains) are\footnote{A set of 
equations that comprise a logical entity, or a system 
of equations that need to be solved simultaneously, are numbered as a block; e.g., {\it N}a, 
{\it N}b, {\it N}c, etc..., where $N$ is an integer.}:

\begin{mathletters}
\beq
\frac{\partial \rho_{\rm n}}{\partial t} + \nabla \cdot (\rho_{\rm n}
\bvec{v}_{\rm n})=0 ,\label{firsteq}
\eeq
\beq
\frac{\partial (\rho_{\rm g_0} + \rho_{\rm g_-} +\rho_{\rm g_+})}{\partial t} + \nabla
\cdot  (\rho_{\rm g_0} \bvec{v}_{\rm g_0} + \rho_{\rm g_-} \bvec{v}_{\rm g_-}
+ \rho_{\rm g_+}
\bvec{v}_{\rm g_+})=0 ,\label{gcont}
\eeq
\beq
\frac{\partial ( \rho_{\rm n} \bvec{v_{\rm n}}) }{\partial t} +
\nabla \cdot (
\rho_{\rm n}  \bvec{v_{\rm n}}  \bvec{v_{\rm n}}) = -\nabla P_{\rm n} +\rho_{\rm n} \bvec{g} 
+ \frac{1}{4\pi} (\nabla \times \bvec{B}) \times
\bvec{B},\label{nforceq}
\eeq
\beq
0 =  -en_{\rm e}( \bvec{E} + \frac{\bvec{v_{\rm e}}}{c} \times \bvec{B}) +
\bvec{F}_{\rm en}, \label{eforceq}
\eeq
\beq
0  =  {\rm e} n_{\rm i}( \bvec{E} + \frac{ \bvec{v_{\rm i}}}{c}  \times \bvec{B}) +
\bvec{F}_{\rm in}, \label{iforceq}
\eeq
\beq
0  =  -{\rm e} n_{\rm g_{-}}( \bvec{E} + \frac{ \bvec{v_{\rm g_{-}}}}{c} \times \bvec{B}) +
\bvec{F}_{\rm g_{-}n} + \bvec{F}_{\rm g_{-}g_{0},inel} \, ,
\label{gmforceq}
\eeq
\beq
0  =  en_{\rm g_{+}}( \bvec{E} + \frac{ \bvec{v_{\rm g_{+}}}}{c} \times \bvec{B})
+\bvec{F}_{\rm g_{+}n} + \bvec{F}_{\rm g_{+}g_{0},inel} \, , 
\label{gpforceq}
\eeq
\beq
0  =  \bvec{F}_{\rm g_{0}n} + \bvec{F}_{\rm g_{0}g_{-},inel}+
\bvec{F}_{\rm g_{0}g_{+},inel} \, ,\label{g0forceq}
\eeq
\beq
\nabla \times \bvec{B} =  \frac{4\pi}{c} \bvec{j},\label{amp}
\eeq
\beq
\bvec{j} =  e( n_{\rm i} \bvec{v}_{\rm i} + n_{\rm g_+} \bvec{v}_{\rm
  g_+} - n_{\rm e} \bvec{v}_{\rm e} - n_{\rm g_-} \bvec{v}_{\rm g_-}
),\label{jdef}
\eeq
\beq
\frac{\partial \bvec{B}}{\partial t}  =  - c (\nabla \times
\bvec{E}),\label{faradayA}
\eeq
\beq
\nabla \cdot \bvec{g}  =  - 4 \pi G \rho_{\rm n},
\eeq
\beq
P_{\rm n}  =  f(\rho_{\rm n},T).\label{lasteq}
\eeq
\end{mathletters}
The quantities $\rho_{\rm s}$, $n_{\rm s}$ and $\bvec{v}_{\rm s}$ denote the mass
density, number density, and velocity of species {\it s}. The subscripts
n, i, e, $\rm g_{-}$, $\rm g_{+}$ and $\rm g_{0}$ refer to the neutrals, ions,
electrons, negatively-charged grains, positively-charged grains, and neutral
grains, respectively. The symbols $\bvec{g}$, $\bvec{E}$ and $\bvec{B}$ denote
the gravitational, electric and magnetic fields, respectively.
The magnetic field satisfies
the condition $\nabla \cdot \bvec{B} = 0$ everywhere at all
times. The definitions and derivations of $\bvec{F}_{s {\rm n}}$ (the per unit volume 
frictional force on species $s$ due to elastic collisions with neutrals) and 
$\bvec{F}_{\gamma \delta,{\rm inel}}$ (the per unit volume force on grain fluid 
$\gamma$ due to the conversion of dust particles of fluid $\delta$ into dust 
particles of fluid $\gamma$) are discussed in detail in \S 3 of \cite{TassisM05a}.

In the force equations of the electrons, ions, and grains, the
acceleration terms have been neglected because of the small inertia of these
species; it was shown by \cite{mpf85} that the plasma reaches a terminal drift 
velocity very fast.
Similarly, the thermal-pressure and gravitational forces (per unit volume) have
been dropped from the force equations of all species other than the neutrals
because they are negligible compared to the electromagnetic and collisional
forces. The inelastic momentum transfer by the electron and ion fluids due to
attachment onto grains and neutralization are negligible compared
to the momentum transfer due to elastic collisions with neutrals, and they have 
been omitted from
the force equations (\ref{eforceq}) and (\ref{iforceq}). 

The equation of state is either isothermal or adiabatic depending on the density.
For densities smaller than a critical density $\rho_{\rm opq}$, the isothermal 
equation of state is used
\beq
P_{\rm n}  =  \rho_{\rm n} C^2,\label{EoSi}
\eeq
where $C = (k_{\rm B} T_{\rm iso}/ \mu m_{\rm H})^{1/2}$ is the isothermal speed 
of sound in the neutrals, $k_{\rm B}$ the Boltzmann constant, 
$\mu$ the mean mass per particle in units of the atomic-hydrogen mass $m_{\rm H}$, and $T_{\rm iso} 
= 10$ K the temperature of the isothermal stages.
For densities greater than $\rho_{opq}$, adiabaticity is assumed and the 
equation of state becomes
\beq
P_{\rm n}  =  \frac{C^2 }{\rho_{\rm opq}^{\gamma - 1}}\rho_{\rm n}^{\gamma},\label{EoSa}
\eeq
where $\gamma$ is the adiabatic exponent.
For simplicity we set
\beq
\gamma = \left\{ \begin{array}{ll}
                   5/3 &{\rm for} \, \, T \leq 200 \,\,{\rm K} \\
                   7/5 &{\rm for} \, \, T > 200 \,\,{\rm K}
                  \end{array}
          \right. 
\eeq
to account for the rotational degrees of freedom of ${\rm H_2}$ 
which only become accessible at high temperatures. 
The temperature $T$ is calculated using Equation (\ref{adiarho}). 

For the early stages of star formation and for the
isothermally infalling gas outside the optically thick inner region, the
temperatures are much smaller than the
grain sublimation temperature ($\approx 1500$ K), and a mass continuity
equation (\ref{gcont}) for the grains (charged + neutral) is appropriate.

Altogether, then, we have a system of 13 equations, (\ref{firsteq})-(\ref{lasteq}), which contain
17 unknowns ($\rho_{\rm n}$, $P$, $\bvec{E}$, $\bvec{B}$, $\bvec{j}$, $\bvec{g}$,
$\bvec{v}_{\rm n}$, $\bvec{v}_{\rm e}$, $\bvec{v}_{\rm i}$, $\bvec{v}_{\rm
  g_-}$, $\bvec{v}_{\rm g_+}$,
$\bvec{v}_{\rm g_0}$, $\rho_{\rm e}$, $\rho_{\rm i}$, $\rho_{\rm g_-}$,
$\rho_{\rm g_+}$,
$\rho_{\rm g_0}$). To close the system, the densities of electrons, ions and charged
grains ($n_{\rm e}$, $n_{\rm i}$, $n_{\rm g-}$, and $n_{\rm g+}$) are
calculated from the equilibrium chemical model [see Appendix A of \cite{TassisM05a}].

\section{Derivation of a Generalized Ohm's Law}\label{ohmslaw}

The evolution of the magnetic field in a fluid depends on the degree of coupling between the 
field and the fluid. If the magnetic field changes in time, an electric field is generated 
in the fluid, as described by Faraday's law of induction\footnote{Note that the sequence of
events described here does not imply causality. An equivalent picture, consistent with 
Maxwell's equations,  would be that the magnetic 
field changes in response to a $\nabla \times \bvec{E}$.}
 (eq. [\ref{faradayA}]). 
This electric field is then capable of driving electric currents through the fluid, which 
in turn generate magnetic fields that tend to oppose the original change. If the conductivity 
of the fluid is infinite,
there are no dissipative losses of the electric currents, 
and the magnetic field is said to be frozen in the fluid. The current generated in response 
to an electric field is described by Ohm's law. A generalized Ohm's law can be derived from 
the force equations of the plasma.

The force equations of all charged species and neutral grains, with the elastic and inelastic 
collision terms written out, are 
\beqar
0 & = & -en_{\rm e}( \bvec{E} + \frac{\bvec{v_{\rm e}}}{c} \times \bvec{B}) +
 \frac{\rho_e}{\tau_{en}}(\bvec{v_{\rm n}}-\bvec{v_{\rm e}}), \label{eforceq2}\\
0 & = & {\rm e} n_{\rm i}( \bvec{E} + \frac{ \bvec{v_{\rm i}}}{c}  \times \bvec{B}) +
 \frac{\rho_i}{\tau_{in}}(\bvec{v_{\rm n}}-\bvec{v_{\rm i}}), \label{iforceq2}\\
0 & = & -{\rm e} n_{\rm g_{-}}( \bvec{E} + \frac{ \bvec{v_{\rm g_{-}}}}{c} \times \bvec{B}) +
\frac{\rho_{\rm g_-}}{\tau_{\rm gn}}(\bvec{v_{\rm n}}-\bvec{v_{\rm g_-}}) \nonumber  \\
    &     & +\frac{\rho_{\rm g_0}}{\tau_-}(\bvec{v_{\rm g_0}}-\bvec{v_{\rm g_-}}),  \label{gmforceq2}\\
0 & = & en_{\rm g_{+}}( \bvec{E} + \frac{ \bvec{v_{\rm g_{+}}}}{c} \times \bvec{B})
+ \frac{\rho_{\rm g_+}}{\tau_{\rm gn}}(\bvec{v_{\rm n}}-\bvec{v_{\rm g_+}}) \nonumber \\
    &    &+\frac{\rho_{\rm g_0}}{\tau_+}(\bvec{v_{\rm g_0}}-\bvec{v_{\rm g_+}}),
\label{gpforceq2}\\
0 &=& \frac{\rho_{\rm g_0}}{\tau_{gn}}(\bvec{v_{\rm n}}-\bvec{v_{\rm g_0}})
+\frac{\rho_{\rm g_0}}{\tau_+}(\bvec{v_{\rm g_+}}-\bvec{v_{\rm g_0}}) \nonumber  \\
    &    &+\frac{\rho_{\rm g_0}}{\tau_-}(\bvec{v_{\rm g_-}}-\bvec{v_{\rm g_0}}),\label{g0forceq2}
\eeqar
where 
\beqar
\tau_- &=& (
\frac{1}{\tau_{\rm g_0e,{\rm inel}}}+
\frac{\rho_{\rm g_-}}{\rho_{\rm g_0}}\frac{1}{\tau_{\rm g_-i,{\rm inel}}} + 
\frac{\rho_{\rm g_-}}{\rho_{\rm g_0}}\frac{1}{\tau_{\rm g_-g_+,{\rm inel}}})^{-1},
\\
\tau_+ &=& (
\frac{1}{\tau_{\rm g_0i,{\rm inel}}}+
\frac{\rho_{\rm g_+}}{\rho_{\rm g_0}}\frac{1}{\tau_{\rm g_+e,{\rm inel}}} + 
\frac{\rho_{\rm g_+}}{\rho_{\rm g_0}}\frac{1}{\tau_{\rm g_+g_-,{\rm inel}}})^{-1}
\eeqar
are the relevant timescales for inelastic collisions between grains.
In particular, $\tau_-$ is the timescale for a neutral grain to be
involved in {\em any} inelastic reaction involving the conversion of
negative grains to neutral grains or {\em vice versa}, and $\tau_+$ is the
timescale for a neutral grain to participate in {\em any} inelastic reaction
involving conversion between positive and neutral grains.

Dividing each of the first four of the above equations by the mass of the
corresponding species, multiplying by $e\tau_{s{\rm n}}$ (where $s$ denotes the
species), multiplying the ion and positive grain equations by -1, and
adding up the resulting equations we obtain 
\beqar
0 &=& -\bvec{E}\left( 
  \frac{e^2n_{\rm e}\tau_{\rm en}}{m_{\rm e}} 
+ \frac{e^2n_{\rm i}\tau_{\rm in}}{m_{\rm i}}
+ \frac{e^2n_{\rm g_{-}}\tau_{\rm gn}}{m_{\rm g}}
+ \frac{e^2n_{\rm g_{+}}\tau_{\rm gn}}{m_{\rm g}}
\right) \nonumber \\
&& - \frac{1}{c}\left(
  \frac{e^2n_{\rm e}\tau_{\rm en}}{m_{\rm e}} \bvec{v_{\rm e}}
+ \frac{e^2n_{\rm i}\tau_{\rm in}}{m_{\rm i}} \bvec{v_{\rm i}}  \right)\times \bvec{B} \nonumber  \\
& & - \frac{1}{c}\left( \frac{e^2n_{\rm g_{-}}\tau_{\rm gn}}{m_{\rm g}} \bvec{v_{\rm g_-}}
+ \frac{e^2n_{\rm g_{+}}\tau_{\rm gn}}{m_{\rm g}}\bvec{v_{\rm g_+}}
\right)\times \bvec{B} \nonumber  \\
&& + e\bvec{v_{\rm n}}\left(n_{\rm e}- n_{\rm i} + n_{\rm g_-} -n_{\rm
g_+} \right)\nonumber \\
&& + e\left(-n_{\rm e}\bvec{v_{\rm e}} + n_{\rm i}\bvec{v_{\rm i}} 
- n_{\rm g_-}\bvec{v_{\rm g_-}} + n_{\rm g_+} \bvec{v_{\rm g_+}}\right)\nonumber \\
&& + e n_{\rm g_0}\left[
\bvec{v_{\rm g_0}}\left( \frac{\tau_{\rm gn}}{\tau_-}-\frac{\tau_{\rm gn}}{\tau_+}\right)
+ \bvec{v_{\rm g_+}}\frac{\tau_{\rm gn}}{\tau_+}
-\bvec{v_{\rm g_-}}\frac{\tau_{\rm gn}}{\tau_-}
\right]. \label{ohm_messy}
\eeqar

In equation (\ref{ohm_messy}), the third term vanishes because of charge
neutrality, while the fourth term is the current density,
$\bvec{j}$. We define the plasma conductivity $\sigma_{\rm p}$ by
\beq\label{def_cond}
\sigma_{\rm p} = 
  \frac{e^2n_{\rm e}\tau_{\rm en}}{m_{\rm e}} 
+ \frac{e^2n_{\rm i}\tau_{\rm in}}{m_{\rm i}}
+ \frac{e^2n_{\rm g_{-}}\tau_{\rm gn}}{m_{\rm g}}
+ \frac{e^2n_{\rm g_{+}}\tau_{\rm gn}}{m_{\rm g}}\,,
\eeq
where the individual contributions to $\sigma_p$ are

\begin{mathletters}
\hfill\begin{tabular}{cc}
$\displaystyle \sigma_{\rm p,e} = \frac{e^2n_{\rm e}\tau_{\rm en}}{m_{\rm e}} $,&
$\displaystyle \sigma_{\rm p,i} = \frac{e^2n_{\rm i}\tau_{\rm in}}{m_{\rm i}}$,
\end{tabular}\hfill(\stepcounter{equation}\theequation,b)
\stepcounter{equation}

\hfill\begin{tabular}{cc}
$\displaystyle \sigma_{\rm p,g_-} = \frac{e^2n_{\rm g_{-}}\tau_{\rm gn}}{m_{\rm g}}$,&
$\displaystyle \sigma_{\rm p,g_+} = \frac{e^2n_{\rm g_{+}}\tau_{\rm gn}}{m_{\rm g}}$.
\end{tabular}\hfill(\stepcounter{equation}\theequation,d)
\stepcounter{equation}
\end{mathletters}

A mean plasma velocity $\bvec{v_{\rm p}}$ can be defined as
\beq\label{def_vp}
\bvec{v_{\rm p}}
= \frac{\sigma_{\rm p,e}}{\sigma_{\rm p}} \bvec{v_{\rm e}}
+ \frac{\sigma_{\rm p,i}}{\sigma_{\rm p}}\bvec{v_{\rm i}}
+\frac{\sigma_{\rm p,g_-}}{\sigma_{\rm p}}\bvec{v_{\rm g_-}}
+ \frac{\sigma_{\rm p,g_+}}{\sigma_{\rm p}}\bvec{v_{\rm g_+}}\,.
\eeq
Then equation (\ref{ohm_messy}) is rewritten as
\beq\label{ohm}
\bvec{E} = - \frac{\bvec{v_{\rm p}}}{c}\times\bvec{B}
+ \frac{\bvec{j}}{\sigma_{\rm p}} + \frac{\bvec{j_0}}{\sigma_{\rm p}}\, ,
\eeq
where 
\beq\label{j0}
\bvec{j_0} = e n_{\rm g_0}\left[
\bvec{v_{\rm g_0}}\left( \frac{\tau_{\rm gn}}{\tau_-}-\frac{\tau_{\rm gn}}{\tau_+}\right)
+ \bvec{v_{\rm g_+}}\frac{\tau_{\rm gn}}{\tau_+}
-\bvec{v_{\rm g_-}}\frac{\tau_{\rm gn}}{\tau_-}\right]\,.
\eeq
The quantity $\bvec{j_0}$ has units of current density and the corresponding term
in the generalized Ohm's law (eq. [\ref{ohm}]) represents the modification 
of this law because of the  
conversion of neutral grains to charged grains and vice versa through inelastic collisions.

Equation (\ref{ohm}) is a different form of the generalized Ohm's law from the usual expression
\begin{equation}\label{usuohm}
\bvec{E_{\rm n}} = \eta_\parallel \bvec{j}_\parallel + 
\eta_\perp \bvec{j}_\perp + \eta_{\rm H} \bvec{j}_\perp \times \hat{\bvec{\rm e}}_B\, 
\end{equation}
\citep{Parks91}, where $\bvec{E_{\rm n}}$ is the electric field in the frame of the neutrals, 
$\parallel$ and $\perp$ denote directions parallel and perpendicular to the magnetic field, 
$\hat{\bvec{\rm e}}_B$ is the unit vector parallel to the magnetic field, and $\eta_\parallel$, 
$\eta_\perp$ and $\eta_{\rm H}$ are the parallel, perpendicular, and Hall resistivities, 
respectively. Equation (\ref{ohm}) can be transformed into equation (\ref{usuohm}), with the 
resistivities appropriately modified to account for the effects of the inelastic couplings 
between grains (see Appendix \ref{conv}).
However, equation (\ref{ohm}) offers several advantages in explicitly expressing the physical 
significance of each process, distinguishing between different mechanisms responsible for 
increasing the mass-to-flux ratio (through Faraday's law of induction) and simplifying the 
numerical implementation of the problem at hand.

First, equation(\ref{ohm}) calculates the electric field in the ``lab'' frame rather than in 
a frame comoving with the neutrals. Although from a mathematical point of view the two 
formulations are equivalent (the electric field in the lab frame is simply obtained by 
subtracting $\bvec{v}_{\rm n}\times \bvec{B} / c$ from the electric field in the frame of 
the neutrals), the interpretation of the results is aided greatly by using the lab frame. 
In this frame, the neutrals are falling toward a center of gravity and the magnetic flux follows,
at an infall speed equal to that of charged species attached to the field lines.
Flux freezing would be a good approximation if the coupling between attached species and the 
neutrals were perfect, in which case the magnetic flux would fall in at the same rate as the 
neutrals. However, under the conditions present in the dense cores of molecular clouds, this 
coupling is imperfect, and the flux is ``left behind'', in a process known as ambipolar 
diffusion. After all species except the underabundant electrons (which interact weakly with 
the rest of the species) have detached, the inward dragging of the magnetic flux with the 
neutral fluid becomes ineffective and its advection speed decreases almost to zero, as shown 
in Paper II. 

If this process is viewed in the frame comoving with the neutrals, it appears that the flux 
has to escape the fluid element by moving outwards. And at late times, when ambipolar 
diffusion is resurrected \citep{DM01}, it would be the responsibility of the extremely few 
electrons to transport this flux outwards -- a process that appears improbable and 
counter-intuitive to some workers, who then suggest alternative mechanisms, such as 
grain drifts \citep{nnu02}. \footnote{\cite{nnu02} claim that \cite{DM01} did not 
properly account for the effect of grains in calculating the Pedersen and Hall conductivities, 
citing equation (28) in \cite{DM01}. In fact, \cite{DM01} did not use equation (28) to 
derive their results. They used equations (7)-(13), which do account for the effects of all 
grain species (although not the effects of inelastic couplings between grain species, which 
were also not considered by \cite{nnu02}, but are explicitly treated here. For the generalized 
Ohm's law including inelastic coupling between grains, see Appendix \ref{conv}). Equation (28) 
in \cite{DM01} is a simplification presented to assist the physical interpretation of the 
results and is not used in any of the numerical simulations.} Such drifts, however, cannot 
possibly be relevant because all charged species except the electrons have detached from 
the field lines by the densities of interest ($\gtrsim 10^{11}$ ${\rm cm^{-3}}$), and the 
contribution of detached species to the drift velocity is insignificant.

Second, equation (\ref{usuohm}) expresses the components of the electric field parallel 
and perpendicular to the field lines, while equation (\ref{ohm}) splits the electric field 
in a more physical way; when it is substituted in Faraday's law of induction 
(eq. [\ref{faradayA}]), it yields
\beq\label{fsplit}
\frac{\partial \bvec{B}}{\partial t} = \nabla \times (\bvec{v}_{\rm p} \times \bvec{B}) - c 
\nabla \times (\frac{\bvec{j}}{\sigma_{\rm p}} + \frac{\bvec{j}_0}{\sigma_{\rm p}}) \,.
\eeq
The first term on the right-hand side represents flux advection\footnote{Here we distinguish 
between flux advection due to the finite velocities of charged species and any flux redistribution 
due to Ohmic dissipation. The first term on the right hand side of equation (\ref{fsplit}) refers only 
to the former effect, not to Ohmic dissipation.}
(including the effect of collisions, i.e., ambipolar diffusion), and the other two 
terms represent flux dissipation due to disruption of currents by collisions (i.e., resistivity). 
The last two terms can be combined into a single term if one defines an ``effective'' 
current $\bvec{j}_{\rm eff} = \bvec{j} + \bvec{j}_0$. In an infinitely conducting medium, 
flux is simply transported inwards with a velocity $\bvec {v}_{\rm p}$, which is the 
conductivity-weighted velocity of the charged species. 

Third, if one uses the electron velocity to advect the magnetic flux rather than the ion velocity, 
then the Hall term in the generalized Ohm's law is absorbed in the ambipolar diffusion term 
(see Mouschovias 1987, eqs. [72a] and [76]). Here, we use a similar approach, which aims at
 simplifying the generalized Ohm's law by taking the advection of the magnetic flux to occur 
with the conductivity-weighted plasma velocity. In this form, the components which are responsible 
for flux dissipation and flux advection are clearly separated. Ohmic losses are represented by 
the $\bvec{j}_{\rm eff}$ term, while flux advection
 is represented by the $\bvec{v}_{\rm p}$ term\footnote{An {\em effective} flux advection 
velocity, representing the apparent velocity of magnetic field lines when Ohmic losses have also 
been accounted for, can also be defined and is equal to $(c/B^2)\bvec{E}\times \bvec{B}$. This effective
flux velocity is our $\bvec{v_{\rm f}}$ (see eq. [\ref{defvf}]) 
and should not be confused with $\bvec{v_{\rm p}}$.}. 
The effect of ambipolar diffusion is then quantified by the difference between $\bvec{v}_{\rm p}$ 
and $\bvec{v}_{\rm n}$. When all species detach (in which case $\bvec{v}_{\rm p}$  becomes equal 
to $\bvec{v}_{\rm n}$), ambipolar diffusion ceases to operate.

\section{Governing Equations in the Thin-Disk Approximation}

Numerical simulations of the ambipolar-diffusion--induced evolution of
axisymmetric, isothermal, molecular clouds have shown that force
balance is rapidly established along field lines and is maintained throughout the evolution
of the model cloud, even during the dynamic phase of contraction  perpendicular to the field 
lines that follows the formation of a thermally and magnetically 
supercritical core \citep{FM92,FM93,DM01}. Following the approach of \cite{CM93},
we take advantage of these results and model the cloud as an axisymmetric thin
(but not infinitesimally thin) disk, with its axis of symmetry aligned
with the $z$-axis of a cylindrical polar
coordinate system $(r,\phi,z)$ and with force balance along the field lines
maintained at all times. The sense in which the thin-disk approximation is
used is that the radial variation of any physical quantity is small over a
distance comparable to the local disk half-thickness.
The upper and lower surfaces of the disk are at $z = +Z(r,t)$ and $z =
-Z(r,t)$, while the radius $R$ of the disk satisfies the condition $R \gg Z$. 
The magnetic field is poloidal and has the configuration described in 
Tassis \& Mouschovias (2005, eqs. [17a] and [17b]). 
The model cloud is embedded in an external medium of constant thermal pressure $P_{\rm ext}$.

For computational simplicity, we reduce the mathematical
dimensionality of the problem by integrating the equations
analytically over $z$, assuming $z$-independence
of all physical quantities inside the model cloud. This is an excellent approximation since
thermal-pressure forces smooth out density gradients over lengthscales smaller than
the critical thermal lengthscale \citep{m91}, which is comparable to $Z(r,t)$. 
A detailed derivation of the thin-disk equations under the assumption of isothermality 
and flux freezing in the ion fluid is given in \cite{CM93}.
Here we extend the formalism to account for breakdown of isothermality and of flux-freezing 
in the electrons.

The continuity equations for the neutrals and the grains in the disk are obtained 
through integration of equations (\ref{firsteq}) and (\ref{gcont}) over $z$, and 
have the form given in Tassis \& Mouschovias (2005, eqs. [18a-c]).
In the isothermal regime,
integration of the $z$-component of the force equation of the neutrals (eq. [\ref{nforceq}]),
with the requirement of force balance along field lines, yields
\begin{mathletters}
\beq\label{thin4}
\rho_{\rm n}(r,t)C^2 = P_{\rm ext} + \frac{\pi}{2}G \sigma_{\rm n}^2(r,t),
\eeq
where $P_{\rm ext}$ is the constant external pressure.
In the adiabatic regime, the pressure becomes
\beqar
P & = & n_{\rm n} k_{\rm B}T \nonumber \\
&=& \rho_{\rm n}\left(\frac{k_{\rm B} T_{\rm iso}}{\mu
  m_{\rm p}}\right)\left(\frac{T}{T_{\rm iso}}\right)\nonumber \\
&=& \rho C^2 \frac{T}{T_{\rm iso}} \,,
\eeqar
so the force balance along the $z-$axis yields in this case
\beq\label{thin4a}
\rho_{\rm n} C^2 \frac{T}{T_{\rm iso}} = P_{\rm ext} + \frac{\pi}{2}G \sigma_{\rm n}^2 .
\eeq
In the isothermal regime, equation (\ref{thin4}) is used to calculate the density
$\rho_{\rm n}(r,t)$ from the column density $\sigma_{\rm n}(r,t)$ and the external pressure. 
In the adiabatic regime, though, equation (\ref{thin4a}) must be solved simultaneously with 
the $\rho_{\rm n}-T$ relation  
\beq\label{adiarho}
\frac{\rho_{\rm n}}{\rho_{\rm opq}} = \left(\frac{T}{T_{\rm opq}}\right)^{\frac{1}{\gamma-1}}
\eeq
to obtain the density $\rho_{\rm n}(r,t)$ and the temperature $T(r,t)$ in the disk.

Integration over $z$ of the $r$-component of the neutral force (eq. [\ref{nforceq}]), 
using the total (thermal-pressure plus Maxwell) stress tensor
\citep{CM93} to combine the thermal-pressure and magnetic-force terms, yields
\beq\label{thin5}
\frac{\partial (\sigma_{\rm n} v_{{\rm n},r})}{\partial t} + \frac{1}{r}\frac{\partial( r 
\sigma_{\rm n} v_{{\rm n},r}^2)}{\partial r}  =   F_{{\rm P},r} + \sigma_{\rm n} g_r + F_{{\rm mag},r} .
\eeq
The expression for the total radial magnetic force (per unit area), 
includes the magnetic tension force (the $B_r$ terms) acting on the surfaces of the disk, and is formally identical to Tassis \& Mouschovias (2005), equation (18g).
The radial component of the thermal-pressure force in the isothermal regime is
\beq
 F_{{\rm P},r} = -C_{\rm eff}^2 \frac{\partial \sigma_{\rm n}}{\partial r}, 
\eeq
where 
\beq
C_{\rm eff}^2 = \frac{\pi}{2}G \sigma_{\rm n}^2 \frac{[3P_{\rm ext} + (\pi/2)G
  \sigma_{\rm n}^2]}{[P_{\rm ext} + (\pi/2)G\sigma_{\rm n}^2]^2} C^2,
\eeq
and accounts for the contribution of the radial force exerted by the external pressure
$P_{\rm ext}$ on the upper and lower surfaces of the disk, since the surfaces
are not horizontal.
In the adiabatic regime, the thermal-pressure force becomes 
(see Appendix \ref{thermadderiv} for derivation) 
\beq
F_{{\rm P},r} = -C_{\rm eff}^2\frac{T}{\tref}\frac{\partial \sigma_{\rm n}}
{\partial r} - C_{\rm new}^2 \sigma_{\rm n}\frac{\partial}{\partial r}
\left(\frac{T}{\tref}\right) ,
\eeq
where
\beq
C_{\rm new}^2 = C^2\frac{\frac{\pi}{2} G \sigma_{\rm n}^2}
{P_{\rm ext}+\frac{\pi}{2}G\sigma_{\rm n}^2}.
\eeq

The $r$-component of the gravitational field calculated from Poisson's equation
for a thin disk has the same form as Tassis \& Mouschovias (2005), equation (18h).

The assumption that the model cloud is embedded in a hot, tenuous, current-free medium 
implies that the magnetic field in the external medium is irrotational
($\nabla \times \bvec{B} = 0$). 
This allows the use of a scalar magnetic potential, $\Psi$, from which the radial component of the 
magnetic field at the top surface of the disk can be obtained (see Tassis \& Mouschovias 2005 eq. [18i]).

In the thin-disk approximation the magnetic field is purely poloidal and from Amp\`{e}re's 
law (eq. [\ref{amp}])
the current density must be purely toroidal, and thus they are orthogonal to each other. 
Consequently, the $\bvec{E}$-field has no component parallel to the magnetic field,
as required by the generalized Ohm's law (eq. [\ref{ohm}]).
We may then define a new quantity, $v_{\rm f}$, which has units of velocity and
can be used to describe the electric field through
\beq\label{defvf}
\bvec{E}=-\frac{\bvec{v_{\rm f}}}{c}\times \bvec{B}\,.
\eeq
Substitution in Faraday's law of induction (eq. [\ref{faradayA}]) yields the evolution equation 
for the magnetic field
\beq\label{Bevol}
\frac{\partial \bvec{B}}{\partial t} = \nabla \times(\bvec{v}_{\rm f} \times \bvec{B}). 
\eeq
In the plane of the disk ($z$ = 0), equation (\ref{Bevol}) becomes
\beq\label{thin8}
\frac{\partial B_{z,{\rm eq}}}{\partial t} + \frac{1}{r}\frac{\partial (r 
B_{z,{\rm eq}} v_{{\rm f},r})}{\partial r}  =  0. 
\eeq
Note that the velocity $v_{\rm f}$ defines a frame of reference in which the magnetic flux is conserved.

The velocities in the $r$ and $\phi$ directions of all species, except the neutrals, 
and $v_{\rm f}$ can be obtained from the solution of the
steady-state force equations of the species and Amp\`{e}re's law 
(eqs. [\ref{eforceq}]-[\ref{g0forceq}] and [\ref{amp}]). The derivation is given in Appendix \ref{velgen}. 
All the {\em radial} velocities have the form 
\beq
v_s = \frac{\Theta_s}{\Theta_s +1}v_{\rm f} + \frac{1}{\Theta_s+1}v_n ,
\eeq
\end{mathletters}
where $\Theta _s$ is the {\it indirect attachment parameter} of species $s$: for $\Theta_s \gg
1$ the $r$-velocity of species $s$ is approximately equal to that of the field lines, and
the species is well attached to the magnetic field, while for $\Theta_s \ll 1$
the species is detached and is falling in with the neutrals ($v_s
\approx v_n$).

\section{The Dimensionless Problem and Method of Solution}

We put the equations in dimensionless form by choosing units 
natural to the model molecular clouds.  The unit of velocity is the isothermal sound speed 
in the neutral fluid, $C$; the units of column density $\sigma_{\rm c,ref}$
and acceleration $2 \pi G \sigma_{\rm c,ref}$ are those of the neutral
column density and gravitational acceleration at the surface on the axis of
symmetry of the reference state from which the initial equilibrium state is calculated. 
The unit of magnetic field is its uniform value in the reference state, $B_{\rm ref}$. 
Although not necessary, we define for computational convenience the unit of temperature 
as the temperature of the isothermal lower density 
parts of the model cloud $\tref =10$ K. The temperature above which the rotational degrees 
of freedom of ${\rm H}_2$ are excited is denoted by $\trot$.

We denote the dimensionless form of the cylindrical polar coordinates
($r$, $z$) by ($\xi$, $\zeta$) and the dimensionless time by $\tau$,
so that the dimensionless thin-disk equations become

\begin{mathletters}
\beq \label{dimless1}
\sigma_{\rm n}(\xi,\tau) =  2\rho_{\rm n}(\xi,\tau)Z(\xi,\tau), 
\eeq
\beq
\frac{\partial \sigma_{\rm n}}{\partial \tau} + \frac{1}{\xi}\frac{\partial (\xi 
\sigma_n v_{{\rm n},\xi})}{\partial \xi}  =  0, 
\eeq
\beq
\frac{\partial (\chi_{\rm g} \sigma_{\rm n})}{\partial \tau} + \frac{1}{\xi}\frac{\partial [\xi 
\sigma_{\rm n} (\chi_{\rm g-} v_{{\rm g-},\xi} + \chi_{\rm g+} v_{{\rm
    g+},\xi} + \chi_{\rm g0}
v_{{\rm g0},\xi})]}{\partial \xi}  =  0, 
\eeq

\beq
\rho_{\rm n}(\xi,\tau) = \left\{ \begin{array}{lr}
              4^{-1}[\tilde{P}_{\rm ext} + \sigma_{\rm n}^2(\xi,\tau)] \, , &  \rho < \rho_{\rm opq} \, ; \\
&\\
              4^{-3/5}\rho_{\rm opq}^{2/5} [\tildpext + 
              \sigma_{\rm n}^2(\xi,\tau)]^{3/5} \, ,  & \\
              & \makebox[0in][r]{$\rho_{\rm opq} < \rho < \rho_{\rm opq} \trot^{3/2} \, ;$} \\
     &\\
              4^{-5/7}\rho_{\rm opq}^{2/7}[\tildpext + 
               \sigma_{\rm n}^2(\xi,\tau)]^{5/7}
                \trot^{-2/7} \, ,   & \\
                & \makebox[0in][r]{$ \rho_{\rm opq} \trot^{3/2} < \rho \, ;$}
                                  \end{array}
          \right.  
\eeq

\beq
T(\xi,\tau) = \left\{ \begin{array}{lr}
              1, \,\,\,\,\,\,\,\,\,\,    &  \rho < \rho_{\rm opq} \, ;\\
              &\\
     4^{-2/5}[\tildpext + \sigma_{\rm n}^2(\xi,\tau)]^{2/5} \rho_{\rm opq}^{\,\,\,-2/5} \, , &\\
           &\makebox[0in][r]{$\rho_{\rm opq} < \rho < \rho_{\rm opq} \trot^{3/2} \, ;$} \\
                                                      &\\
4^{-2/7}[\tildpext + \sigma_{\rm n}^2(\xi,\tau)]^{2/7} 
\rho_{\rm opq}^{\,\,\,-2/7}\trot^{2/7} \, , &\\
              & \makebox[0in][r]{$\rho_{\rm opq} \trot^{3/2} < \rho \, ;$}
                                  \end{array}
          \right. 
\eeq

\beq
\frac{\partial (\sigma_{\rm n} v_{{\rm n},\xi})}{\partial \tau} + \frac{1}{\xi}\frac{\partial( \xi 
\sigma_{\rm n} v_{{\rm n},\xi}^2)}{\partial \xi}  =  F_{{\rm P},\xi} + \sigma_{\rm n} g_{\xi} + F_{{\rm mag},\xi} \, ; 
\label{dimlessforc}
\eeq

\beq
F_{{\rm P},\xi} = \left\{ \begin{array}{ll}
            \displaystyle  -\tilde{C}_{\rm eff}^2 \frac{\partial \sigma_{\rm n}}{\partial \xi} \, ,  &  \rho < \rho_{\rm opq} \, ;\\
            \displaystyle   -\tilde{C}_{\rm eff}^2 T(\xi) \frac{\partial \sigma_{\rm n}}{\partial \xi} 
               -\tilde{C}_{\rm new} \sigma_{\rm n}(\xi) \frac{\partial \trot}{\partial \xi} \, , & \rho_{\rm opq} < \rho \, ;
                                  \end{array}
          \right. 
\eeq

\beq
\tilde{C}_{\rm eff}^2 = \sigma_{\rm n}^2 \frac{(3\tilde{P}_{\rm ext} +
  \sigma_{\rm n}^2)}{(\tilde{P}_{\rm ext} + \sigma_{\rm n}^2)^2},
\eeq

\beq
\tilde{C}_{\rm new}^2 = \frac{\sigma_{\rm n}^2}{\tilde{P}_{\rm ext} + \sigma_{\rm n}^2},
\eeq

\beqar
&&F_{{\rm mag},\xi} = \frac{1}{\mu_{d,c0}^2} \left\{ B_{\zeta,{\rm eq}}\left( B_{\xi,Z}-Z\frac{\partial
  B_{\zeta,{\rm eq}}}{\partial \xi}\right) \right. \nonumber \\ 
&& \left.+ \frac{1}{2}
\frac{\partial Z}{\partial \xi} \left[B_{\xi,Z}^2
+ 2B_{\zeta,{\rm eq}}\left(B_{\xi,Z}\frac{\partial Z}{\partial \xi}\right)+\left(B_{\xi,Z}\frac{\partial
  Z}{\partial \xi}\right)^2\right] \right\}, \nonumber \\ 
\eeqar
\beq
g_{\rm \xi}(\xi) = \int_{0}^{\infty}d\xi' \xi' \sigma_{\rm n}(\xi')\mathcal{M}(\xi,\xi'),
\eeq
\beq
\mathcal{M}(\xi,\xi') = \frac{2}{\pi}\frac{d}{d\xi}\left[ \frac{1}{\xi_>}K\left(
  \frac{\xi_<}{\xi_>}\right)  \right],
\eeq
\beq
B_{\xi,Z}(\xi) = - \int_{0}^{\infty}d\xi' \xi' [B_{\zeta,{\rm eq}}(\xi')-B_{\rm ref}]\mathcal{M}(\xi,\xi'),
\eeq
\beq
\frac{\partial B_{\zeta,{\rm eq}}}{\partial \tau} + \frac{1}{\xi}\frac{\partial (\xi 
B_{\zeta,{\rm eq}} v_{{\rm f},\xi})}{\partial \xi}  =  0. 
\eeq
\end{mathletters}
%%%%%%%%%%%%%%%%%%%%%%%%%%%%%%%%%%%%%%%%%%%%%%%%%%%%%%%%%%%%%%%%%%%%%%%%%%%%%%%%%%%%%%%

\noindent
Equations (\ref{dimless1}-n) and the equations for the velocities of 
all species (Appendix \ref{velgen}) comprise the dimensionless system to be
solved numerically. The dependent variables are now
dimensionless but in this section only, for ease in identification, we
have retained the same symbols as for their dimensional counterparts.

The free parameters of the problem are: 
the initial central mass-to-flux ratio
$\mu_{d,c0}=(2\pi G^{1/2}\sigma_{c,ref})/B_{ref}$ in units of its
critical value for gravitational collapse in disk geometry,
$1/(2\pi G^{1/2})$ \citep{nn78};
the constant external
pressure relative to the central gravitational ``pressure'' along the
axis of symmetry of
the reference state $\tilde{P}_{\rm ext} \equiv P_{\rm ext}/
(\frac{\pi}{2}G\sigma_{c,ref})$;
and the density $n_{\rm opq}$, above which the equation of state
becomes adiabatic.

The reaction rates of the chemical network and the value of the radius of the
dust grains are given in Appendix A of \cite{TassisM05a}. [The thermal (Jeans) length does
  not appear explicitly in the equations because of
the choice of the units we are using: The unit of length
$C^2 / 2 \pi G \sigma_{\rm c,ref}$ is proportional to the Jeans length.]

A control-volume formulation is employed for solving numerically the dimensionless
equations that govern the evolution of the model cloud. The numerical method consists of an 
implicit time integrator; a conservative upwind advective difference scheme
that incorporates the second-order accurate monotonicity algorithm of \cite{vL79}; a 
second-order difference approximation of forces inside a mass
zone; an integral approximation of the gravitational and magnetic fields; and
an adaptive mesh capable of resolving the smallest natural lengthscale
(i.e., thermal lengthscale $\lambda_{T,crit}$). 
Using convergence tests, we found that 85 zones in the radial direction are adequate 
to follow the evolution of the core to the densities of interest with an accuracy better
than 1\%.  
The algebraic equations
for the equilibrium abundances of charged species are solved at each time
step. A detailed description of the numerical methods is given by \cite{mmc94}.

In order to follow shocks that might arise during the evolution, we change the 
originally central-difference scheme for the calculation of spatial derivatives to a 
second-order accurate one-sided difference scheme \citep{pm83}. In addition,
tensor artificial viscosity \citep{TW79} is introduced. The expression for the 
artificial viscosity term that is added to the RHS of the radial force equation 
(\ref{dimlessforc}) is given in \cite{TassisM05a}. 

Our initial and boundary conditions are the same as those discussed in detail in 
\S 5 of \cite{TassisM05a}. 

\section{Discussion}

We have presented the formulation of the problem of the ambipolar-diffusion--controlled 
formation and evolution of magnetically supercritical molecular cloud cores inside 
initially magnetically subcritical parent clouds, and the core collapse to high densities, 
past the detachment of ions from the magnetic field and into the opaque regime, in which 
we approximated the treatment of radiation trapping using an adiabatic equation of state. 
We have used the six-fluid MHD equations to describe the model cloud. The magnetic flux 
was not assumed to be frozen in any of the charged species, in order to allow for the 
study of the possible detachment of the lightest charged species, the electrons, which 
would signify the complete decoupling of the magnetic field from the matter. 

We have reduced the numerical complexity of the problem by integrating the equations 
analytically along field lines, and we have used force-balance along the same direction 
to calculate the thickness of the formed magnetic disk. 

We have derived a new expression for the generalized Ohm's law. It includes the effects 
of inelastic couplings between grains, which not only modify the expression for the 
contribution of each charged species to the total conductivity, but also account for 
the contribution of {\em neutral} grains to the conductivity (because neutral grains 
become charged part of the time). Moreover, this form of Ohm's law, when substituted 
in Faraday's law of induction, explicitly distinguishes between advection of the 
magnetic flux by charged species and Ohmic dissipation, which involves disruption 
of currents by collisions. 

In two companion papers, we present the solution of the problem for a model cloud 
with fiducial parameters (Paper II) and a detailed parameter study (Paper III).

\acknowledgements{KT would like to thank  Arieh K\"{o}nigl, Matt Kunz, and Vasiliki Pavlidou for 
useful discussions, 
and Glenn Ciolek for providing the base version of the code used in this work. The work of KT was 
supported in part by the University of Illinois through a Dissertation Completion Fellowship. This 
research was partially supported by a grant from the American Astronomical Society and NSF grants 
AST 02-06216 and AST 02-39759.}

%%%%%%%%%% Appendix
\appendix

\section{Generalized Ohm's Law: Equivalence of Different Formulations}
\label{conv}

In this Appendix we present the derivation of a generalized Ohm's law (in its usual directional 
formulation) from the six-fluid MHD force equations, accounting explicitly for the inelastic 
coupling between grain species. The  inelastic grain coupling not only modifies the perpendicular, 
parallel, and Hall resistivities, but also introduces a second dissipative term associated with 
the  neutral grain fluid. 

In \S \, \ref{ohmslaw} we presented an alternative formulation of the generalized Ohm's law, 
which is written in a maximally simplified functional form by determining the appropriate, 
conductivity-weighted plasma velocity, with which the magnetic flux is being advected. In 
this Appendix we discuss how the directional formulation of the generalized Ohm's law is 
related to this maximally simplified formulation.
In addition, we present a general solution of the velocities of all charged species as a 
function of the velocity of the neutrals and an effective magnetic-flux velocity.

\subsection{General Solution for the Velocities of Charged Species and Neutral Grains}

We adopt a local cartesian coordinate system such that the $x-$axis is parallel to 
$\bvec{j}\times \bvec{B}$, the $y-$axis is parallel to 
$(\bvec{j}\times \bvec{B})\times \bvec{B}$, and the $z-$axis parallel to $\bvec{B}$. 
We can then write 
\beq
\begin{array}{cccc}
\bvec{B} = B\hat{\bvec{e}}_z\,, & \bvec{j}=j_\perp\hat{\bvec{e}}_y+j_\parallel\hat{\bvec{e}}_z\,,
& \bvec{j}\times\bvec{B} = j_\perp B  \hat{\bvec{e}}_x\,,&
(\bvec{j}\times\bvec{B})\times \bvec{B} = -j_\perp B^2 \hat{\bvec{e}}_y \, ,
\end{array}
\eeq
where $\hat{\bvec{e}}_x$, $\hat{\bvec{e}}_y$ and $\hat{\bvec{e}}_z$ are the three orthonormal 
unit vectors. In addition, we define an effective magnetic-flux velocity $\bvec{v}_{\rm f}$ by
\beq
-\frac{1}{c}\bvec{v}_{\rm f}\times\bvec{B} = \bvec{E}_\perp.
\eeq
The physical meaning of this velocity is that it represents the velocity of the boundary of 
a constant-flux surface. The effects responsible for the motion of this boundary can be 
motion of the fluid element, flux redistribution between fluid elements, or flux dissipation 
due to Ohmic losses. With these definitions, we can write the force equations 
(\ref{eforceq})-(\ref{g0forceq}) in component form in the $x-y$ plane as 

\begin{mathletters}
\beqar
0&=& \oeten(v_{{\rm f} y}- v_{{\rm e} y}) + v_{{\rm n} x} -v_{{\rm e} x} \label{myeqer}\\
0&=& \oeten(v_{{\rm e} x} - v_{{\rm f} x}) +v_{{\rm n} y} -v_{{\rm e} y} \label{myeqef}\\
0&=& \oitin(v_{{\rm i} y} -v_{{\rm f} y}) + v_{{\rm n} x} - v_{{\rm i} x} \label{myeqir}\\
0&=& \oitin(v_{{\rm f} x}-v_{{\rm i} x}) +v_{{\rm n} y}-v_{{\rm i} y} \label{myeqif}\\
0&=& \ogtgn (v_{{\rm f} y}-v_{{\rm g-} y})+v_{{\rm n} x} -v_{{\rm g-} x} +
\rmm(v_{{\rm g0} x}-v_{{\rm g-} x}) \label{myeqmr}\\
0&=& \ogtgn(v_{{\rm g-} x} - v_{{\rm f} x}) +v_{{\rm n} y}-v_{{\rm g-} y} + 
\rmm(v_{{\rm g0} y}-v_{{\rm g-} y})\label{myeqmf}\\
0&=& \ogtgn(v_{{\rm g+} y} - v_{{\rm f} y}) + v_{{\rm n} x} - v_{{\rm g+} x} + 
\rpp(v_{{\rm g0} x}-v_{{\rm g+} x})\label{myeqpr}\\
0&=& \ogtgn(v_{{\rm f} x}-v_{{\rm g+} x}) +v_{{\rm n} y} -v_{{\rm g+} y} + 
\rpp(v_{{\rm g0} y}-v_{{\rm g+} y})\label{myeqpf}\\
0&=& v_{{\rm n} x} - v_{{\rm g0} x} + \frac{\tgn}{\tm}(v_{{\rm g-} x}-v_{{\rm g0} x}) + \frac{\tgn}{\tp}(v_{{\rm g+} x}-v_{{\rm g0} x})\label{myeq0r}\\
0&=& v_{{\rm n} y}- v_{{\rm g0} y} + \frac{\tgn}{\tm}(v_{{\rm g-} y}-v_{{\rm g0} y}) + \frac{\tgn}{\tp}(v_{{\rm g+} y}-v_{{\rm g0} y})\label{myeq0f}
\eeqar
\end{mathletters}

Once the $x-$ and $y-$ components of the velocity of the neutrals and the magnetic-flux 
velocity have been specified, the $x-$ and $y-$ velocities of the other species can be 
found by solving the linear, algebraic system of the force equations of the species (in 
steady state). In matrix form, the system of
equations (\ref{myeqer}) - (\ref{myeq0f}) is
\beq
{\cal A}\bvec{V}= v_{{\rm n} x} \bvec{C^{{\rm n} x}} + v_{{\rm f} x} \bvec{C^{{\rm f}x}}
+v_{{\rm n} y} \bvec{C^{{\rm n} y}} + v_{{\rm f} y} \bvec{C^{{\rm f}y}}
\,,
\eeq
where 
${\cal A}$ is the matrix of the coefficients of the unknowns 
{\arraycolsep=0in
\beqar
{\cal A} &=&
\left[\begin{array}{cccccccccc}\vspace{3mm}
-1 & 0 & 0 & 0& 0& -\oeten & 0 & 0 & 0 &0 \\ \vspace{3mm}
\oeten & 0 & 0 & 0 & 0 &-1 & 0 & 0&0& 0 \\ \vspace{3mm}
0&-1 & 0 & 0& 0& 0&\oitin& 0& 0& 0 \\ \vspace{3mm}
0&-\oitin & 0 & 0 & 0 & 0& -1 & 0& 0& 0\\ \vspace{3mm}
0&0& -1-r_- & 0& r_- &0 &0 &-\ogtgn & 0&0\\ \vspace{3mm}
0&0&\ogtgn & 0&0&0&0& -1-r_- & 0&r_-\\ \vspace{3mm}
0&0&0& -1-r_+ & r_+&0&0&0&\ogtgn& 0 \\ \vspace{3mm}
0&0&0&-\ogtgn&0& 0&0&0&-1-r_+ & r_+ \\ \vspace{3mm}
0&0&\displaystyle \frac{\tgn}{\tm}& \displaystyle \frac{\tgn}{\tp} & 
\displaystyle -\frac{\tgn}{\tz} & 0&0 & 0&0&0\\ \vspace{3mm}
0&0&0&0&0& 
0&0&\displaystyle \frac{\tgn}{\tm} & \displaystyle \frac{\tgn}{\tp} & 
\displaystyle -\frac{\tgn}{\tz}
\end{array}\right]\nonumber \\
\eeqar
}

with 

\begin{mathletters}
\hfill\begin{tabular}{ccc}
$\displaystyle \tau_0 = \left( \frac{1}{\tau_{gn}}+\frac{1}{\tau_-}+\frac{1}{\tau_+}
\right)^{-1}$, &
$\displaystyle r_- = \rmm$, &
$\displaystyle r_+ = \rpp$, 
\end{tabular}\hfill(\stepcounter{equation}\theequation,b,c)
\stepcounter{equation}
\stepcounter{equation}
\end{mathletters}

and 

\beq
\begin{array}{ccccc}
\bvec{V} = \left[ \begin{array}{c}
v_{{\rm e} x} \\ v_{{\rm i} x} \\ v_{{\rm g-} x} \\ v_{{\rm g+} x} 
\\ v_{{\rm g0} x} \\ v_{{\rm e} y} \\ v_{{\rm i} y} \\ v_{{\rm g-} y}  
\\ v_{{\rm g+} y} \\ v_{{\rm g0} y}
\end{array} \right],&
\bvec{C^{{\rm n}x}} = \left[ \begin{array}{r}
-1\\0\\-1 \\0\\-1\\0\\-1 \\0\\-1\\0
\end{array} \right],&
 \bvec{C^{{\rm f}x}} = \left[ \begin{array}{c}
0\\ \oeten \\0 \\-\oitin\\0\\\ogtgn\\0 \\-\ogtgn\\0\\0
\end{array} \right],&
\bvec{C^{{\rm n}y}} = \left[ \begin{array}{r}
0\\-1\\0\\-1 \\0\\-1\\0\\-1 \\0\\-1
\end{array} \right],&
 \bvec{C^{{\rm f}y}} = \left[ \begin{array}{c}
 -\oeten \\0 \\ \oitin\\0\\-\ogtgn\\0 \\ \ogtgn\\0\\0\\0
\end{array} \right]
\end{array}\,.
\eeq

We use Cramer's method to solve the above system. We define 
\beq
D=det[{\cal A}]\,.
\eeq
In addition, we use the notation $D^{{\rm n}x}_{sx}$ to represent the
determinant of $\cal A$ where the column corresponding to the $x-$component of species 
$s$ has been substituted by $\bvec{C^{{\rm n}x}}$, the notation $D^{{\rm n}x}_{sy}$ 
for the determinant derived from $D$ if the $y-$component of species $s$ column is 
replaced by $\bvec{C^{{\rm n}x}}$, and similarly for the other $\bvec{C}$ vectors. 
Then, the solution of the system is, 

\beq
v_{sx} = \frac{D^{{\rm n}x}_{sx}}{D}v_{{\rm n} x} 
+\frac{D^{{\rm f}x}_{sx}}{D} v_{{\rm f}x}
+ \frac{D^{{\rm n}y}_{sx}}{D}v_{{\rm n} y} 
+\frac{D^{{\rm f}y}_{sx}}{D} v_{{\rm f}y}
\eeq
for the $x-$components of the velocities, and 
\beq
v_{sy} = \frac{D^{{\rm n}x}_{sy}}{D}v_{{\rm n} x} 
+\frac{D^{{\rm f}x}_{sy}}{D} v_{{\rm f}x}
+ \frac{D^{{\rm n}y}_{sy}}{D}v_{{\rm n} y}
+\frac{D^{{\rm f}y}_{sy}}{D} v_{{\rm f}y}
\eeq
for the $y-$components. 

The determinants defined above satisfy the relations
\beq\label{properties}
D^{{\rm n}i}_{sj}+D^{{\rm f}i}_{sj} = D \delta_{ij} \, ,
\eeq
where $i$ and $j$ can take one of the values $x$, $y$ and $\delta_{ij}$ 
is the Kronecker delta. 
These properties can be easily proved as follows. The 
determinant $D^{{\rm n}i}_{sj}+D^{{\rm f}i}_{sj}$ 
is the same as the determinant which we get if we substitute the
$sj$-column of $D$ 
by $\bvec{C^{{\rm f}i}}+\bvec{C^{{\rm n}i}}$. If from this 
column we then subtract the 3 other $j$-columns, we get either the 
determinant of the coefficients $D$ (if $i=j$), or a determinant which is identically 
zero (if $i\ne j$).
We define
\beq
\Theta_{sx} = \frac{D^{{\rm f}x}_{sx}}{D^{{\rm n}x}_{sx}}
\eeq
\beq
\Theta_{sy} = \frac{D^{{\rm f}y}_{sy}}{D^{{\rm n}y}_{sy}}
\eeq
\beq
\Lambda_{sx} = \frac{D^{{\rm f}y}_{sx}}{D} = - \frac{D^{{\rm n}y}_{sx}}{D}
\eeq
\beq
\Lambda_{sy} = \frac{D^{{\rm f}x}_{sy}}{D} = - \frac{D^{{\rm n}x}_{sy}}{D} \, .
\eeq
With these definitions and the properties (\ref{properties}), the solution can be re-written as 
\beq\label{xs}
v_{sx} = \frac{1}{\Theta_{sx}+1}v_{{\rm n}x}
+ \frac{\Theta_{sx}}{\Theta_{sx}+1} v_{{\rm f}x}
+ \Lambda_{sx}(v_{{\rm f}y}-v_{{\rm n}y})
\eeq
for the $x-$components and
\beq\label{ys}
v_{sy} = \frac{1}{\Theta_{sy}+1}v_{{\rm n}y}
+ \frac{\Theta_{sy}}{\Theta_{sy}+1}v_{{\rm f}y}
+ \Lambda_{sy}(v_{{\rm f}x}-v_{{\rm n}x})
\eeq
for the $y-$components.

\subsection{Drift of Magnetic Flux Relative to the Neutrals as a Function of the Current Density}

Let us define $\bvec{w}_{s} = \bvec{v}_{s} - \bvec{v}_{\rm n}$ and 
$\bvec{w}_{\rm f} = \bvec{v}_{\rm f} - \bvec{v}_{\rm n}$. Adding up the force equations 
for all species (including the neutral grains) we obtain the plasma force equation, 
\beq \label{plasma_ap}
\frac{B^2}{c^2}\sigma_{\rm p} \sum_{s} \frac{1}{(\omega_{s}\tau_{s{\rm n}})^2}\frac{\sigma_{s}}
{\sigma_{\rm p}}\bvec{w}_{s} = \frac{1}{c} \, \bvec{j} \times \bvec{B}
\,.
\eeq
Taking the $x-$ and $y-$components of equation (\ref{plasma_ap}) we find
\beq \label{plasmax}
\frac{B}{c}\sigma_{\rm p} \sum_{s}\frac{1}{(\omega_{s}\tau_{s{\rm n}})^2}\frac{\sigma_{s}}
{\sigma_{\rm p}}w_{sx} = j_\perp 
\, ,
\eeq
and
\beq \label{plasmay}
\sum_{s}\frac{1}{(\omega_{s}\tau_{s{\rm n}})^2}\frac{\sigma_{s}}{\sigma_{\rm p}}w_{ sy} = 0
\,.
\eeq
In addition, equations (\ref{xs}) and (\ref{ys}) give:
\begin{mathletters}
\beq \label{wsx}
w_{sx} = \frac{\Theta_{sx}}{\Theta_{sx}+1}w_{{\rm f}x} + \Lambda_{sx} w_{{\rm f}y}
\eeq
and
\beq\label{wsy}
w_{sy} = \frac{\Theta_{sy}}{\Theta_{sy}+1}w_{{\rm f}y} + \Lambda_{sy} w_{{\rm f}x} \, .
\eeq
\end{mathletters}
Hence, equations (\ref{plasmax}) and (\ref{plasmay}) can be rewritten as
\begin{mathletters}
\beq \label{plasmaxb}
w_{{\rm f}x} \sum_{s}\frac{1}{(\omega_{s}\tau_{s{\rm n}})^2}
\frac{\sigma_{s}}{\sigma_{\rm p}}\frac{\Theta_{sx}}{\Theta_{sx}+1} +
w_{{\rm f}y} \sum_{s}\frac{1}{(\omega_{s}\tau_{s {\rm n}})^2}
\frac{\sigma_{s}}{\sigma_{\rm p}} \Lambda_{sx} = 
\frac{c}{B}\frac{j_\perp}{\sigma_{\rm p}} 
\, ,
\eeq
and
\beq \label{plasmayb}
w_{{\rm f}y} \sum_{s}\frac{1}{(\omega_{s}\tau_{s{\rm n}})^2}
\frac{\sigma_{s}}{\sigma_{\rm p}} \Lambda_{sy} + 
w_{{\rm f}y} \sum_{s} \frac{1}{(\omega_{s}\tau_{s{\rm n}})^2}
\frac{\sigma_{s}}{\sigma_{\rm p}}\frac{\Theta_{sy}}{\Theta_{sy}+1} = 0 
\,.
\eeq
\end{mathletters}
For convenience and brevity, we define the coefficients of the above equations to be 
\beq
\begin{array}{cc}
\displaystyle \theta_{i} = \sum_{s} \frac{1}{(\omega_{s}\tau_{s{\rm n}})^2}
\frac{\sigma_{s}}{\sigma_{\rm p}}\frac{\Theta_{si}}{\Theta_{si}+1}\,, & 
\displaystyle \lambda_{i} = \sum_{s}\frac{1}{(\omega_{s}\tau_{s{\rm n}})^2}\frac{\sigma_{s}}
{\sigma_{\rm p}} \Lambda_{si}
\end{array}
\eeq
with $i$ again being either $x$ or $y$. Then, the solution of equations (\ref{plasmaxb}) 
and (\ref{plasmayb}) with respect to $w_{{\rm f}x}$ and $w_{{\rm f}y}$ is 
\beq\label{solution}
\begin{array}{cc}
\displaystyle w_{{\rm f}x} = \frac{c}{B}\frac{j_\perp}{\sigma_{\rm p}}
\frac{\theta_{y}}{\theta_{x}\theta_{y} - \lambda_{x}\lambda_{y}} \,,
& 
\displaystyle w_{{\rm f}y} = -\frac{c}{B}\frac{j_\perp}{\sigma_{\rm p}}
\frac{\lambda_{y}}{\theta_{x}\theta_{y} - \lambda_{x}\lambda_{y}} \, .
\end{array}
\eeq

\subsection{Equivalence of Equation (\ref{ohm}) and (\ref{usuohm}) for Ohm's Law}

Adding $(\bvec{v}_{\rm n}/c)\times \bvec{B}$ to both sides of equation (\ref{ohm}) we obtain 
\beq\label{ohmintermed}
\bvec{E}_{\rm n} = \bvec{E} + \frac{1}{c} \bvec{v_{\rm n}} \times \bvec{B}= 
- \frac{\bvec{v}_{\rm p} - \bvec{v}_{\rm n}}{c}\times\bvec{B}
+ \frac{\bvec{j}}{\sigma_{\rm p}} + \frac{\bvec{j_0}}{\sigma_{\rm p}} \, ,
\eeq
where $\bvec{E}_{\rm n}$ is the electric field in a frame comoving with the neutrals.
We rewrite the first term in the right-hand side of equation (\ref{ohmintermed}) in terms 
of $w_{{\rm f}x}$ and $w_{{\rm f}y}$, using the definition of $\bvec{v}_{\rm p}$ (eq. 
[\ref{def_vp}]) to introduce the drift velocities $w$ and equations (\ref{wsx}) and (\ref{wsy}) 
to eliminate $w_{cx}$ and $w_{cy}$ in favor of $w_{{\rm f}x}$ and $w_{{\rm f}y}$:
\beqar
\frac{\bvec{v}_{\rm p} - \bvec{v}_{\rm n}}{c}\times\bvec{B} &=& 
\frac{1}{c}\left[\left(\sum_{k}\frac{\sigma_{k}}{\sigma_{\rm p}}\bvec{v}_{k}\right)-\bvec{v}_{\rm n}\right]\times \bvec{B} \nonumber \\
 &=& \frac{1}{c}\sum_{k}\frac{\sigma_{k}}{\sigma_{\rm p}}\left(\bvec{v}_{k} -\bvec{v}_{\rm n}\right)\times \bvec{B}  \nonumber \\
&=& \frac{1}{c}\sum_{k}\frac{\sigma_{k}}{\sigma_{\rm p}}\bvec{w}_{k}\times \bvec{B} \nonumber \\
&=& \hat{\bvec{e}}_x \frac{B}{c} \sum_{k}\frac{\sigma_{k}}{\sigma_{\rm p}}w_{ky} - \hat{\bvec{e}}_y \frac{B}{c} \sum_{k}\frac{\sigma_{k}}{\sigma_{\rm p}}w_{kx} \nonumber \\
&=& \hat{\bvec{e}}_x \frac{B}{c} \left[w_{{\rm f}y}\sum_{k}\frac{\sigma_{k}}{\sigma_{\rm p}}\frac{\Theta_{ky}}{\Theta_{ky}+1}
+ w_{{\rm f}x}\sum_{k}\frac{\sigma_{k}}{\sigma_{\rm p}}\Lambda_{ky}\right]
\nonumber \\
&& - \hat{\bvec{e}}_y \frac{B}{c} \left[w_{{\rm f}x}\sum_{k}\frac{\sigma_{k}}{\sigma_{\rm p}}
 \frac{\Theta_{kx}}{\Theta_{kx}+1}
+ w_{{\rm f}y}\sum_{k}\frac{\sigma_{k}}{\sigma_{\rm p}}\Lambda_{kx}\right]\,, 
\eeqar
Note that $k$ counts over charged species only, while $s$ counts over all species (including 
neutral grains). Now substituting $w_{{\rm f}x}$ and $w_{{\rm f}y}$ using the solution 
(\ref{solution}), and defining the sums above to be 
\beq
\begin{array}{cc}
\displaystyle \Theta_{i} = \sum_{k} \frac{\Theta_{ki}}{\Theta_{ki}+1}
\frac{\sigma_{k}}{\sigma_{\rm p}}\,, & 
\displaystyle \Lambda_{i} = \sum_{k}\Lambda_{ki}\frac{\sigma_{k}}{\sigma_{\rm p}}\,,
\end{array}
\eeq
where $i=x,y$, we obtain

\beqar
\frac{\bvec{v}_{\rm p} - \bvec{v}_{\rm n}}{c}\times\bvec{B} &=& 
\hat{\bvec{e}}_x \frac{j_\perp}{\sigma_{\rm p}}\left[
-\Theta_{y} \frac{\lambda_{y}}{\theta_{x}\theta_{y} - \lambda_{x}\lambda_{y}} + \Lambda_{y} 
\frac{\theta_{y}}{\theta_{x}\theta_{y} - \lambda_{x}\lambda_{y}}\right] \nonumber \\
&& -\hat{\bvec{e}}_y \frac{j_\perp}{\sigma_{\rm p}}\left[
\Theta_{x} \frac{\theta_{y}}{\theta_{x}\theta_{y} - \lambda_{x}\lambda_{y}} - \Lambda_{x} 
\frac{\lambda_{y}}{\theta_{x}\theta_{y} - \lambda_{x}\lambda_{y}}
\right] \nonumber \\
&=& \frac{1}{\sigma_{\rm p}}\bvec{j}_\perp\times \hat{\bvec{e}}_z \left[
-\Theta_{y} \frac{\lambda_{y}}{\theta_{x}\theta_{y} - \lambda_{x}\lambda_{y}} + \Lambda_{y} 
\frac{\theta_{y}}{\theta_{x}\theta_{y} - \lambda_{x}\lambda_{y}}\right] \nonumber \\
&&- \frac{1}{\sigma_{\rm p}}\bvec{j}_\perp \left[
\Theta_{x} \frac{\theta_{y}}{\theta_{x}\theta_{y} - \lambda_{x}\lambda_{y}} - \Lambda_{x} 
\frac{\lambda_{y}}{\theta_{x}\theta_{y} - \lambda_{x}\lambda_{y}}
\right] 
\eeqar
or, defining
\beq
\eta_{\rm adv,H} = \frac{1}{\sigma_{\rm p}}\left[
-\Theta_{y} \frac{\lambda_{y}}{\theta_{x}\theta_{y} - \lambda_{x}\lambda_{y}} + \Lambda_{y} 
\frac{\theta_{y}}{\theta_{x}\theta_{y} - \lambda_{x}\lambda_{y}}\right]
\eeq
and
\beq
\eta_{\rm adv,\perp} = \frac{1}{\sigma_{\rm p}}\left[
-\Theta_{x} \frac{\theta_{y}}{\theta_{x}\theta_{y} - \lambda_{x}\lambda_{y}} + \Lambda_{x} 
\frac{\lambda_{y}}{\theta_{x}\theta_{y} - \lambda_{x}\lambda_{y}}
\right] 
\eeq
we can write the advection term as 
\beq \label{term1}
\frac{\bvec{v}_{\rm p} - \bvec{v}_{\rm n}}{c}\times\bvec{B} =
\eta_{\rm adv,H}\bvec{j}_\perp\times \hat{\bvec{e}}_z + 
\eta_{\rm adv,\perp} \bvec{j}_\perp \, .
\eeq
The other two components of equation (\ref{ohmintermed}) can be analyzed as 
\beq \label{term2}
\frac{\bvec{j}}{\sigma_{\rm p}} = \frac{\bvec{j}_\parallel}{\sigma_{\rm p}}
+ \frac{\bvec{j}_\perp}{\sigma_{\rm p}}
\eeq
and, with the definition $\sigma_{\rm g0}=e^2n_{\rm g0}\tau_{\rm gn}/m_{\rm g}$,
\beqar
\frac{\bvec{j}_0}{\sigma_{\rm p}} &=&
 \frac{\bvec{j}_{0\parallel}}{\sigma_{\rm p}}+ 
\frac{1}{\sigma_{\rm p}}\frac{B}{c}\frac{1}{\omega_{\rm g}\tau_{\rm gn}}\sigma_{\rm g0}
\left[\bvec{w}_{\rm g0}\left(\frac{\tau_{\rm gn}}{\tau_-}-\frac{\tau_{\rm gn}}
{\tau_+}\right)+\bvec{w}_{\rm g+}\frac{\tau_{\rm gn}}{\tau_+}-\bvec{w}_{\rm g-}
\frac{\tau_{\rm gn}}{\tau_-}\right] \nonumber \\
&=& \hat{\bvec{e}}_x \frac{\sigma_{\rm g0}}{\sigma_{\rm p}}\frac{B}{c}\frac{1}
{\omega_{\rm g}\tau_{\rm gn}} \left\{w_{{\rm f}x} \left[
\left(\frac{\tau_{\rm gn}}{\tau_-}-\frac{\tau_{\rm gn}}{\tau_+}\right) 
\frac{\Theta_{{\rm g0}x}}{\Theta_{{\rm g0}x}+1} 
+ \frac{\tau_{\rm gn}}{\tau_+}\frac{\Theta_{{\rm g+}x}}{\Theta_{{\rm g+}x}+1}
- \frac{\tau_{\rm gn}}{\tau_-}\frac{\Theta_{{\rm g-}x}}{\Theta_{{\rm g-}x}+1}
\right]+ \right.\nonumber \\
&& \left. w_{{\rm f}y}\left[
\left(\frac{\tau_{\rm gn}}{\tau_-}-\frac{\tau_{\rm gn}}{\tau_+}\right)Z_{{\rm g0}x}
+ \frac{\tau_{\rm gn}}{\tau_+}Z_{{\rm g+}x}
- \frac{\tau_{\rm gn}}{\tau_-}Z_{{\rm g-}x}
\right]
\right\}\nonumber \\
&& + \hat{\bvec{e}}_y \frac{\sigma_{\rm g0}}{\sigma_{\rm p}}\frac{B}{c}\frac{1}
{\omega_{\rm g}\tau_{\rm gn}} \left\{w_{{\rm f}y} \left[
\left(\frac{\tau_{\rm gn}}{\tau_-}-\frac{\tau_{\rm gn}}{\tau_+}\right) 
\frac{\Theta_{{\rm g0}y}}{\Theta_{{\rm g0}y}+1} 
+ \frac{\tau_{\rm gn}}{\tau_+}\frac{\Theta_{{\rm g+}y}}{\Theta_{{\rm g+}y}+1}
- \frac{\tau_{\rm gn}}{\tau_-}\frac{\Theta_{{\rm g-}y}}{\Theta_{{\rm g-}y}+1}
\right]+ \right. \nonumber \\
&& \left. w_{{\rm f}x}\left[
\left(\frac{\tau_{\rm gn}}{\tau_-}-\frac{\tau_{\rm gn}}{\tau_+}\right)\Lambda_{{\rm g0}y}
+ \frac{\tau_{\rm gn}}{\tau_+}\Lambda_{{\rm g+}y}
- \frac{\tau_{\rm gn}}{\tau_-}\Lambda_{{\rm g-}y}
\right] 
\right\} \, .
\eeqar
If we now define 
\beqar
\eta_{\rm 0H} &=& \frac{1}{\sigma_{\rm p}}\frac{\sigma_{\rm g0}}{\sigma_{\rm p}}
\frac{1}{\omega_{\rm g}\tau_{\rm gn}} \left\{ \frac{\theta_{y}}
{\theta_{x}\theta_{y}-\lambda_{x}\lambda_{y}} \left[\left(\frac{\tau_{\rm gn}}
{\tau_-}-\frac{\tau_{\rm gn}}{\tau_+}\right) \frac{\Theta_{{\rm g0}x}}{\Theta_{{\rm g0}x}+1} 
+ \frac{\tau_{\rm gn}}{\tau_+}\frac{\Theta_{{\rm g+}x}}{\Theta_{{\rm g+}x}+1}
- \frac{\tau_{\rm gn}}{\tau_-}\frac{\Theta_{{\rm g-}x}}{\Theta_{{\rm g-}x}+1}
\right] \right. \nonumber \\
&&\left. - \frac{\lambda_{y}}{\theta_{x}\theta_{y}-\lambda_{x}\lambda_{y}}
\left[
\left(\frac{\tau_{\rm gn}}{\tau_-}-\frac{\tau_{\rm gn}}{\tau_+}\right)\Lambda_{{\rm g0}x}
+ \frac{\tau_{\rm gn}}{\tau_+}\Lambda_{{\rm g+}x}
- \frac{\tau_{\rm gn}}{\tau_-}\Lambda_{{\rm g-}x}
\right]
\right\}\nonumber\\
\eeqar
and 
\beqar
\eta_{\rm 0 \perp} &=&  \frac{1}{\sigma_{\rm p}}\frac{\sigma_{\rm g0}}{\sigma_{\rm p}}
\frac{1}{\omega_{\rm g}\tau_{\rm gn}} \left\{ \frac{\theta_{y}}
{\theta_{x}\theta_{y}-\lambda_{x}\lambda_{y}} \left[\left(\frac{\tau_{\rm gn}}
{\tau_-}-\frac{\tau_{\rm gn}}{\tau_+}\right)\Lambda_{{\rm g0}y}
+ \frac{\tau_{\rm gn}}{\tau_+}\Lambda_{{\rm g+}y}
- \frac{\tau_{\rm gn}}{\tau_-}\Lambda_{{\rm g-}y}
\right] \right. \nonumber\\
&&  \left. - \frac{\lambda_{y}}{\theta_{x}\theta_{y}-\lambda_{x}\lambda_{y}}
\left[ \left(\frac{\tau_{\rm gn}}{\tau_-}-\frac{\tau_{\rm gn}}{\tau_+}\right) 
\frac{\Theta_{{\rm g0}y}}{\Theta_{{\rm g0}y}+1} 
+ \frac{\tau_{\rm gn}}{\tau_+}\frac{\Theta_{{\rm g+}y}}{\Theta_{{\rm g+}y}+1}
- \frac{\tau_{\rm gn}}{\tau_-}\frac{\Theta_{{\rm g-}y}}{\Theta_{{\rm g-}y}+1}
\right] \right\}\nonumber \, .\\ 
\eeqar
we can rewrite the $\bvec{j}_0$ term as
\beq \label{term3}
\frac{\bvec{j}_0}{\sigma_{\rm p}} = \frac{\bvec{j}_{0\parallel}}
{\sigma_{\rm p}}+ \eta_{\rm 0H} \bvec{j}_\perp\times \hat{\bvec{e}_{\rm z}}
+ \eta_{\rm 0\perp} \bvec{j}_\perp\,.
\eeq

Substituting equations (\ref{term1}), (\ref{term2}), and (\ref{term3}) in equation 
(\ref{ohmintermed}), we find that
\beq
\bvec{E}_{\rm n} = \frac{1}{\sigma_{\rm p}}\left(\bvec{j}_\parallel+\bvec{j}_{0 \parallel}\right) 
+ \left(\frac{1}{\sigma_{\rm p}}+\eta_{\rm adv,\perp} + \eta_{\rm 0,\perp}\right) \bvec{j}_\perp
+\left( \eta_{\rm adv,H}+\eta_{\rm 0H}\right)\bvec{j}_\perp\times \hat{\bvec{e}_{\rm z}} \, .
\eeq
Note that the inelastic collisions between grains enter this expression both explicitly 
(through the new term $\bvec{j}_{0\parallel}$ and the new contributions to the perpendicular 
and Hall conductivities $\eta_{0\perp}$ and $\eta_{0\rm H}$) as well as implicitly, by altering 
the expressions for $\eta_{\rm adv, \perp}$ and $\eta_{\rm adv,H}$.

In this Appendix we showed how a generalized Ohm's law can be derived from the six-fluid 
MHD force equations with the inelastic coupling between grains taken into account. We 
discussed two different formulations of the generalized Ohm's law. In the first case, 
the electric field is separated directionally, into a component parallel to the magnetic 
field, a component parallel to the component of the current which is perpendicular to 
the magnetic field, and a third component perpendicular to both magnetic field and current. 
In the second case, the contributions to the electric field are separated into an advection 
term and a term quantifying the contribution of each charge species to the Ohmic losses 
when substituted into Faraday's law of induction. The formulations are mathematically 
equivalent (and we have explicitly showed how one transforms into the other). Each offers 
different advantages for physical understanding and interpretation of the formalism and 
for application in numerical simulations. 

We have presented new formule for all relevant resistivities for the case that inelastic 
grain couplings are taken into account. We have shown that the inelastic grain coupling 
not only modifies the perpendicular, parallel and Hall resistivities, but also introduces 
an extra dissipative term associated with the resistance to the motion of the neutral 
grain fluid. Finally, we have presented a general solution of the velocities of all 
charged species as a function of the velocity of the neutrals and an effective magnetic-flux 
velocity. The expressions presented in this paper can readily be used in non-ideal MHD 
codes for the study of weakly ionized astrophysical fluids. Depending on the problem and 
the code, one or the other form of the generalized Ohm's law can be used.

%%%%%%%%%%%%%%%%%%%%%% Appendix B %%%%%%%%%%%%%%%%%%%%%%%%%%%%%%%%%%%%%%
\section{Derivation of the Thermal - Pressure Force in the Thin-Disk Approximation}
\label{thermadderiv}

We define a volume element $V$ of the disk as the cylinder-like volume that 
an arbitrary surface $A$ in the midplane ($z=0$) sweeps 
when it is translated vertically until it reaches the top $z=Z(x,y)$ and bottom $z=-Z(x,y)$ 
surfaces. The total force exerted on that 
volume element $V$ in the $i$ direction is
\beq
F_i = \int_V \nabla \cdot \tens{T} dV,
\eeq
or, using the divergence theorem 
\beq
F_i = \oint_S dS \, \, \tens{T} \cdot \hat{n},
\eeq
where $\tens{T}$ is the thermal-pressure stress tensor, $S$ is the bounding surface of the 
volume $V$, and $\hat{n}$ the local unit vector normal to the surface.
This force can be expressed in terms of the area $A$ on the midplane,
perpendicular to the field lines:
\beq
F = \int_A dA \, F_{A}.
\eeq
The integrand $F_{A}$ is the force per unit area on the midplane of the disk.
We need to determine $F_{A}$.

We split the bounding surface $S$ into three surfaces: top, bottom, and side. Then the 
contribution of each one to $F$ can be calculated separately, and 
\beq
F = F_{\rm top} + F_{\rm bottom} + F_{\rm side}.
\eeq
The unit normal vectors of the top and bottom surfaces of the disk are

\begin{mathletters}
\hfill\begin{tabular}{cc}
$\displaystyle \hat{n}_{\rm top}=\frac{\hat{z}-\nabla_\parallel Z}{\sqrt{1+(\nabla_\parallel Z)^2}}$, &
$\displaystyle \hat{n}_{\rm bottom}=\frac{-\hat{z}-\nabla_\parallel Z}{\sqrt{1+(\nabla_\parallel Z)^2}}$,
\end{tabular}\hfill(\stepcounter{equation}\theequation,b)
\stepcounter{equation}
\end{mathletters}

\noindent
where $\nabla_\parallel$ is the derivative in the direction parallel to the midplane, 
perpendicular to the field lines inside the disk.
The surface element of the top/bottom surface is related to the surface element of the 
midplane area $A$ as
\beq
dS_{\rm top/bottom} = \sqrt{1+(\nabla_\parallel Z)^2} \, dA.
\eeq
The intersection of the side surface with the midplane is a closed curve $l$. 
Thus the surface element of the side surface is 
\beq
dS_{\rm side}= dl \cdot dz,
\eeq
and the normal unit vector is always on the midplane.
So
\beqar
F_{\rm top} &=& \int_{S_{\rm top}} dS_{\rm top} \hat{n}_{\rm top} \cdot \tens{T} (z=Z+\epsilon) \nonumber \\
            &=& \int dA (\hat{z}-\nabla_\parallel Z) \cdot \tens{T} (z=Z+\epsilon).
\eeqar 
Similarly,
\beq
F_{\rm bottom} = \int dA (-\hat{z}-\nabla_\parallel Z) \cdot \tens{T} (z=-Z-\epsilon),
\eeq 
and 
\beq
F_{\rm side} = \oint dl \int_{-Z}^{+Z} dz \, \hat{n}_{\parallel} \cdot \tens{T} \, \,.
\eeq 

The contribution of each surface to the thermal-pressure force is
\beqar
F_{\rm top} &=& -\int dA (\hat{z}-\nabla_\parallel Z) P_{\rm ext} ,\\
F_{\rm bottom} &=& -\int dA (-\hat{z}-\nabla_\parallel Z) P_{\rm ext} ,\\
F_{\rm side} &=& -\oint dl \int_{-Z}^{+Z} dz \, \hat{n}_{\parallel} P ,\nonumber \\
             &=& -\int dA \, \nabla_\parallel 2 Z P. 
\eeqar
The total thermal-pressure force in the adiabatic regime is 
\beq \label{B14}
F_P = -\int dA \, \nabla_\parallel 
\left( \sigma_{\rm n} C^2\frac{T}{\tref} -2P_{\rm ext}Z\right)\,.
\eeq
Force balance in the $z-$direction implies
\beq 
\rho_{\rm n}C^2\frac{T}{\tref} = 
P_{\rm ext} + \frac{\pi G\sigma_{\rm n}^2}{2},
\eeq
so that 
\beq
Z = \frac{\sigma_{\rm n}}{2\rho_{\rm n}} = 
\frac{\sigma_{\rm n}C^2\frac{T}{\tref}}
{2\rho_{\rm n}C^2\frac{T}{\tref}}
 = \frac{\sigma_{\rm n}C^2\frac{T}{\tref}}
{2P_{\rm ext}+ \pi G \sigma_{\rm n}^2}\,.
\eeq
Therefore, equations (\ref{B14}) becomes
\beqar
F_P &=& - \int dA \, \nabla _\parallel 
\left(\sigma_{\rm n } C^2 \frac{T}{\tref}
-2P_{\rm ext}\frac{\sigma_{\rm n}C^2\frac{T}{\tref}}
{2P_{\rm ext}+ \pi G \sigma_{\rm n}^2}\right) \nonumber \\ 
&=&  - \int dA \, C^2 
\frac{\frac{\pi}{2} G \sigma_{\rm n}^2}
{P_{\rm ext}+ \frac{\pi}{2} G \sigma_{\rm n}^2}
\left[ \sigma_{\rm n}\nabla _\parallel 
\frac{T}{\tref} +
\frac{T}{\tref}
\frac{3P_{\rm ext}+ \frac{\pi}{2} G \sigma_{\rm n}^2 }
{P_{\rm ext}+ \frac{\pi}{2} G \sigma_{\rm n}^2}
\nabla _\parallel \sigma_{\rm n}
\right] \nonumber  \, \, .
\eeqar
We define
\beq
C_{\rm eff}^2 = C^2\,\,
\frac{\frac{\pi}{2} G \sigma_{\rm n}^2
\left(3P_{\rm ext}+\frac{\pi}{2}G\sigma _{\rm n}^2\right)}
{\left(P_{\rm ext}+\frac{\pi}{2}G\sigma_{\rm n}^2\right)^2}
\eeq
and
\beq
C_{\rm new}^2 = C^2\frac{\frac{\pi}{2} G \sigma_{\rm n}^2}
{P_{\rm ext}+\frac{\pi}{2}G\sigma_{\rm n}^2}.
\eeq
Then the pressure force is written in a compact form as
\beq
F_P = -\int dA \left[
C_{\rm new}^2 \sigma_{\rm n}\nabla _\parallel \frac{T}{\tref}
+C_{\rm eff}^2 \frac{T}{\tref} \nabla_\parallel \sigma_{\rm n}
\right]\,.
\eeq
In the $r-$direction, \[\nabla_\parallel \sigma_{\rm n}
=\frac{\partial \sigma_{\rm n}}{\partial r}\] and 
\[\nabla_\parallel \frac{T}{\tref}
=\frac{\partial}{\partial r}\left(\frac{T}{\tref}\right),\]
so that the radial thermal-pressure force per unit area
in the adiabatic regime is 
\beq \label{presfr}
F_{{\rm P},r} = -C_{\rm eff}^2\frac{T}{\tref}\frac{\partial \sigma_{\rm n}}
{\partial r} - C_{\rm new}^2 \sigma_{\rm n}\frac{\partial}{\partial r}
\left(\frac{T}{\tref}\right).
\eeq
Clearly, if $T = T_{iso}$ the second term vanishes, and equation (\ref{presfr}) reduces 
to the expression (28a) of \cite{CM93} for the thermal-pressure force. If $T > T_{iso}$ , 
the temperature gradient contributes to the thermal-pressure force.

%%%%%%% Appendix C %%%%%%%%%%%%%%%%%%%%%%%%%%%%%%%%%%%%%%%%%%%%%%%%%%%%%%%%%%

\section{Species Velocities} 
\label{velgen}

Once the velocity of the neutrals, $v_{\rm n}$, has been specified, the
velocities of the other species can be found by solving the linear,
algebraic system of the force equations of the species (in steady state)
and Amp\`{e}re's law (eqs. [\ref{eforceq2}] - [\ref{g0forceq2}] and [\ref{amp}],
where $\bvec{j}$ in eq. [\ref{amp}] is given by eq. [\ref{jdef}]).

We use equation (\ref{defvf}) to eliminate the electric field from
equations (\ref{eforceq2}) - (\ref{gpforceq2}):
\beqar
0&=&-\frac{en_{\rm e}}{c}(\bvec{v_{\rm e}}-\bvec{v_{\rm f}})\times \bvec{B} 
+ \frac{\rho_{\rm e}}{\tau_{\rm en}}(\bvec{v_{\rm n}}-\bvec{v_{\rm e}})\label{esimple}\\
0&=&\frac{en_{\rm i}}{c}(\bvec{v_{\rm i}}-\bvec{v_{\rm f}})\times \bvec{B} 
+ \frac{\rho_{\rm i}}{\tau_{\rm in}}(\bvec{v_{\rm n}}-\bvec{v_{\rm i}})\label{isimple}\\
0&=&-\frac{en_{\rm g-}}{c}(\bvec{v_{\rm g-}}-\bvec{v_{\rm f}})\times \bvec{B} 
+ \frac{\rho_{\rm g-}}{\tau_{\rm gn}}(\bvec{v_{\rm n}}-\bvec{v_{\rm g-}})
+\frac{\rho_{\rm g0}}{\tau_-}(\bvec{v_{\rm g0}}-\bvec{v_{\rm g-}})
\label{msimple}
\\
0&=&\frac{en_{\rm g+}}{c}(\bvec{v_{\rm g+}}-\bvec{v_{\rm f}})\times \bvec{B} 
+ \frac{\rho_{\rm g+}}{\tau_{\rm gn}}(\bvec{v_{\rm n}}-\bvec{v_{\rm g+}})
+\frac{\rho_{\rm g0}}{\tau_+}(\bvec{v_{\rm g0}}-\bvec{v_{\rm g+}}).
\label{psimple}
\eeqar
In addition, we eliminate $\bvec{E}$ (using eq. [\ref{defvf}]) and
$\bvec{j}$ (using eq. [\ref{amp}]) from equation (\ref{ohm}), the generalized Ohm's law, to obtain:
\beqar
-\frac{\bvec{v_{\rm f}}}{c}\times \bvec{B} &=& 
- \frac{1}{c}  
\left(\frac{\sigma_{\rm p,e}}{\sigma_{\rm p}}\bvec{v_{\rm e}}
+\frac{\sigma_{\rm p,i}}{\sigma_{\rm p}}\bvec{v_{\rm i}}
+\frac{\sigma_{\rm p,g_-}}{\sigma_{\rm p}}\bvec{v_{\rm g_-}}
+\frac{\sigma_{\rm p,g_+}}{\sigma_{\rm p}}\bvec{v_{\rm g_+}}
\right)\times\bvec{B}\nonumber \\
&&+ \frac{c\bvec{\nabla}\times \bvec{B}}{4\pi\sigma_{\rm p}} 
+ \frac{en_{\rm g_0}}{\sigma_{\rm p}}\left[
\left(\frac{\tau_{\rm gn}}{\tau_-}-\frac{\tau_{\rm gn}}{\tau_+}\right)\bvec{v_{\rm g_0}}+
\frac{\tau_{\rm gn}}{\tau_+}\bvec{v_{\rm g_+}}
-\frac{\tau_{\rm gn}}{\tau_-}\bvec{v_{\rm g_-}}
\right]\,\,,\label{ohmvf}
\eeqar
where we have also used equations (\ref{def_vp}) and (\ref{j0}) to eliminate 
$\bvec{v_{\rm p}}$ and $\bvec{j_0}$ in terms of the velocities of the species.

Equations (\ref{esimple}) - (\ref{ohmvf}) and the neutral-grain force equation, 
\beq\label{nsimple}
0= \frac{\rho_{g0}}{\tau_{gn}}(\bvec{v_{n}}-\bvec{v_{g0}})
+\frac{\rho_{g0}}{\tau_+}(\bvec{v_{g+}}-\bvec{v_{g0}})
+\frac{\rho_{g0}}{\tau_-}(\bvec{v_{g-}}-\bvec{v_{g0}})\,\,,
\eeq
in component form constitute a $12\times 12$ linear system of equations, with unknowns 
the $r-$ and $\phi-$ components of the velocities of electrons, ions, 
neutral, positive and negative grains, and the $r-$ and $\phi-$
components of
$\bvec{v_f}$. 

We now integrate equations (\ref{esimple}) - (\ref{nsimple}) 
in the $z$-direction using the
one-zone approximation (no variation of quantities with $z$ inside the
disk) and decompose them into $r-$ and $\phi-$ components
(assuming we have no bulk rotation of the disk, $v_{n\phi}=0$). Note
that {\it ratios} of densities, $\rho$, 
equal ratios of column densities, $\sigma$, and the former rather than
the latter are used whenever such ratios appear. 
Also, note that the magnetic field contribution in all equations
other than the generalized Ohm's law (through
the gyrofrequencies of the species) only comes from the $B_z$
component. Although the integration over $z$ will in principle bring
out the contribution of $B_r$ from the upper and lower surfaces of the
disc, such contribution does not actually appear in the
$\phi-$components of the equations since it would be multiplied by 
a difference of $z-$ velocities which identically vanishes (there are no drifts in the $z-$direction).
In the case of the integrated $r-$component of the generalized Ohm's law, $B_r$
contributes through $F_{\rm mag}$.
The resulting $12 \times 12$ linear system (with unknowns the velocities $\vi, \vif, \ve, \vef, \vgm,
 \vgmf, \vgp, \vgpf, \vgo, \vgof, v_{\rm f},v_{\rm f \phi}$) is:
\begin{mathletters}
\beqar
0&=& \oeten(v_{\rm f \phi}-\vef) + \vn -\ve \label{eqer}\\
0&=& \oeten(\ve - v_{\rm f}) -\vef \label{eqef}\\
0&=& \oitin(\vif-v_{\rm f \phi}) + \vn -\vi \label{eqir}\\
0&=& \oitin(v_{\rm f}-\vi) -\vif \label{eqif}\\
0&=& \ogtgn (v_{\rm f \phi}-\vgmf)+\vn -\vgm +\rmm(\vgo-\vgm) \label{eqmr}\\
0&=& \ogtgn(\vgm - v_{\rm f}) -\vgmf + \rmm(\vgof-\vgmf)\label{eqmf}\\
0&=& \ogtgn(\vgpf - v_{\rm f \phi}) + \vn - \vgp + \rpp(\vgo-\vgp)\label{eqpr}\\
0&=& \ogtgn(v_{\rm f}-\vgp) -\vgpf + \rpp(\vgof-\vgpf)\label{eqpf}\\
0&=& \vn - \vgo + \frac{\tgn}{\tm}(\vgm-\vgo) + \frac{\tgn}{\tp}(\vgp-\vgo)\label{eq0r}\\
0&=& - \vgof + \frac{\tgn}{\tm}(\vgmf-\vgof) + \frac{\tgn}{\tp}(\vgpf-\vgof)\label{eq0f}\\
v_{\rm f\phi}&=&   
\frac{\sigma_{\rm p,e}}{\sigma_{\rm p}}v_{\rm e\phi}
+\frac{\sigma_{\rm p,i}}{\sigma_{\rm p}}v_{\rm i\phi}
+ \frac{\sigma_{\rm p,g_-}}{\sigma_{\rm p}}v_{\rm g_-\phi}
+ \frac{\sigma_{\rm p,g_+}}{\sigma_{\rm p}}v_{\rm g_+\phi}\nonumber\\
&&
-\frac{\sigma_{\rm p,g_0}}{\sigma_{\rm p}}
\left[\left(\frac{1}{\omega_{\rm g }\tau_-}-\frac{1}
{\omega_{\rm g}\tau_+}\right)v_{\rm g_0}+
\frac{1}{\omega_{\rm g}\tau_+}v_{\rm g_+}
-\frac{1}{\omega_{\rm g}\tau_-}v_{\rm g_-}
\right]\label{ohmr}\\
v_{\rm f}&=&   
\frac{\sigma_{\rm p,e}}{\sigma_{\rm p}}v_{\rm e}
+\frac{\sigma_{\rm p,i}}{\sigma_{\rm p}}v_{\rm i}
+ \frac{\sigma_{\rm p,g_-}}{\sigma_{\rm p}}v_{\rm g_-}
+ \frac{\sigma_{\rm p,g_+}}{\sigma_{\rm p}}v_{\rm g_+}\nonumber\\
&&+\frac{F_{\rm mag} \tau_{\rm pn}}{\sigma_{\rm n}}
+\frac{\sigma_{\rm p,g_0}}{\sigma_{\rm p}}
\left[\left(\frac{1}{\omega_{\rm g }\tau_-}-\frac{1}
{\omega_{\rm g}\tau_+}\right)v_{\rm g_0 \phi}+
\frac{1}{\omega_{\rm g}\tau_+}v_{\rm g_+ \phi}
-\frac{1}{\omega_{\rm g}\tau_-}v_{\rm g_- \phi}
\right] , \label{ohmf}
\eeqar
\end{mathletters}
where $\sigma_{\rm p, g_0}$ has units of conductivity and is defined as
\beq
\sigma_{\rm p, g_0} = \frac{e^2 n_{\rm g_0}\tau_{\rm gn}}{m_{\rm g}} \, ,
\eeq
and $\tau_{\rm pn}$ is a properly averaged collisional timescale 
between the plasma (charged species) and the neutrals and is defined by
\beq
\frac{1}{\tau_{\rm pn}}=
\frac{\rho_{\rm e}}{\rho_{\rm n}}
\frac{(\omega_{\rm e}\tau_{\rm en})^2}{\tau_{\rm en}}
+\frac{\rho_{\rm i}}{\rho_{\rm n}}
\frac{(\omega_{\rm i}\tau_{\rm in})^2}{\tau_{\rm in}}
+\frac{\rho_{\rm g_+}}{\rho_{\rm n}}
\frac{(\omega_{\rm g}\tau_{\rm gn})^2}{\tau_{\rm gn}}
+\frac{\rho_{\rm g_-}}{\rho_{\rm n}}
\frac{(\omega_{\rm g}\tau_{\rm gn})^2}{\tau_{\rm gn}}\,.
\eeq
Note that in equation (\ref{ohmf}), $\sigma_n$ is the column density of the neutrals,
 not a conductivity. 

We solve the system of equations (\ref{eqer}) - (\ref{ohmr}) in terms 
of $\vn$ and $v_{\rm f}$, and then substitute all velocities in equation 
(\ref{ohmf}) to find $v_{\rm f}$ in terms of $\vn$. In matrix form, the system of
equations (\ref{eqer}) - (\ref{ohmr}) is
\beq
{\cal A}\bvec{V}= \vn \bvec{C^n} + v_{\rm f} \bvec{C^f}\,,
\eeq
where 
${\cal A}$ is the matrix of the coefficients of the unknowns 
{\arraycolsep=0in
\beqar
{\cal A} &=&
\left[\begin{array}{ccccccccccc}\vspace{3mm}
-1 & 0 & 0 & 0& 0& -\oeten & 0 & 0 & 0 & 0 & \oeten \\ \vspace{3mm}
\oeten & 0 & 0 & 0 & 0 &-1 & 0 & 0& 0& 0& 0 \\ \vspace{3mm}
0&-1 & 0 & 0& 0& 0&\oitin& 0& 0& 0& -\oitin \\ \vspace{3mm}
0&-\oitin & 0 & 0 & 0 & 0& -1 & 0& 0& 0& 0\\ \vspace{3mm}
0&0& -1-r_- & 0& r_- &0 &0 &-\ogtgn & 0&0& \ogtgn\\ \vspace{3mm}
0&0&\ogtgn & 0&0&0&0& -1-r_- & 0&r_-&0\\ \vspace{3mm}
0&0&0& -1-r_+ & r_+&0&0&0&\ogtgn& 0 & -\ogtgn \\ \vspace{3mm}
0&0&0&-\ogtgn&0& 0&0&0&-1-r_+ & r_+ & 0 \\ \vspace{3mm}
0&0& \displaystyle \frac{\tgn}{\tm}& \displaystyle \frac{\tgn}{\tp} & 
 \displaystyle -\frac{\tgn}{\tz} & 0&0 & 0&0&0&0\\ \vspace{3mm}
0&0&0&0&0& 
0&0& \displaystyle \frac{\tgn}{\tm} & \displaystyle \frac{\tgn}{\tp} & 
\displaystyle -\frac{\tgn}{\tz}& 0 \\0&0&\varphi_-&-\varphi_+&\varphi_0& 
\displaystyle \frac{\sigma_{\rm p,e}}{\sigma_{\rm p}}&
\displaystyle \frac{\sigma_{\rm p,i}}{\sigma_{\rm p}}&
\displaystyle \frac{\sigma_{\rm p,g_-}}{\sigma_{\rm p}}& 
\displaystyle \frac{\sigma_{\rm p,g+}}{\sigma_{\rm p}}
&0 & -1 \\
\end{array}\right]\nonumber \\
\eeqar
}
and

\begin{mathletters}
\hfill\begin{tabular}{ccc}
$\displaystyle \tau_0 = \left( \frac{1}{\tau_{gn}}+\frac{1}{\tau_-}+\frac{1}{\tau_+}
\right)^{-1}$, &
$\displaystyle r_- = \rmm$, &
$\displaystyle r_+ = \rpp$, 
\end{tabular}\hfill(\stepcounter{equation}\theequation,b,c)
\stepcounter{equation}
\stepcounter{equation}

\hfill\begin{tabular}{ccc}
$\displaystyle \varphi_- = \frac{\sigma_{\rm p,g_0}}{\sigma_{\rm p} \omega_{\rm g}\tm}$, & 
$\displaystyle \varphi_+ = \frac{\sigma_{\rm p,g_0}}{\sigma_{\rm p} \omega_{\rm g}\tp}$, &
$\displaystyle \varphi_0 = \varphi_+-\varphi_-$,
\end{tabular}\hfill(\stepcounter{equation}\theequation,e,f)
\end{mathletters}
 
\beq
\begin{array}{ccc}
\bvec{V} = \left[ \begin{array}{c}
\ve \\ \vi \\\vgm\\\vgp\\\vgo\\ \vef \\ \vif \\ \vgmf\\ \vgpf \\ \vgof \\ v_{\rm f \phi}
\end{array} \right] \,\,,&
\bvec{C^n} = \left[ \begin{array}{r}
-1\\0\\-1 \\0\\-1\\0\\-1 \\0\\-1\\0 \\0
\end{array} \right] \,\,,&
 \bvec{C^f} = \left[ \begin{array}{c}
0\\ \oeten \\0 \\-\oitin\\0\\\ogtgn\\0 \\-\ogtgn\\0\\0 \\0
\end{array} \right]
\end{array}\,.
\eeq

We use Cramer's method to solve the above system. We define 
\beq
D=det[{\cal A}]\, ,
\eeq
and use the notation $D^n_s$ to represent the
determinant of $\cal A$ in which the column corresponding to $s$ (one of 
the $\rm 
e, e_{\phi}, i, i_{\phi}, g_-, g_{- \phi}, g_+, g_{+ \phi}, g_0, g_{0 \phi}, f_{\phi}$) has been 
substituted by $\bvec{C^n}$. Similarly, $D^f_s$ is the
determinant of $\cal A$ with the column $s$ having been replaced by
$\bvec{C^f}$. Then, the solution of the system is 

\begin{mathletters}
\hfill\begin{tabular}{ll}
$\displaystyle \ve = \frac{D^n_e}{D}\vn + \frac{D^f_e}{D}\vf$,&
$\displaystyle \vef = \frac{D^n_{e\phi}}{D}\vn + \frac{D^f_{e\phi}}{D}\vf$,
\end{tabular}\hfill(\stepcounter{equation}\theequation,b)
\stepcounter{equation}

\hfill\begin{tabular}{ll}
$\displaystyle \vi = \frac{D^n_i}{D}\vn + \frac{D^f_i}{D}\vf$,&
$\displaystyle \vif = \frac{D^n_{i\phi}}{D}\vn + \frac{D^f_{i\phi}}{D}\vf$,
\end{tabular}\hfill(\stepcounter{equation}\theequation,b)
\stepcounter{equation}

\hfill\begin{tabular}{ll}
$\displaystyle \vgm = \frac{D^n_{g-}}{D}\vn + \frac{D^f_{g-}}{D}\vf$,&
$\displaystyle \vgmf = \frac{D^n_{g-\phi}}{D}\vn + \frac{D^f_{g-\phi}}{D}\vf$,
\end{tabular}\hfill(\stepcounter{equation}\theequation,d)
\stepcounter{equation}

\hfill\begin{tabular}{ll}
$\displaystyle \vgp = \frac{D^n_{g+}}{D}\vn + \frac{D^f_{g+}}{D}\vf $,&
$\displaystyle \vgpf = \frac{D^n_{g+\phi}}{D}\vn + \frac{D^f_{g+\phi}}{D}\vf $,
\end{tabular}\hfill(\stepcounter{equation}\theequation,f)
\stepcounter{equation}

\hfill\begin{tabular}{ll}
$\displaystyle \vgo = \frac{D^n_{g0}}{D}\vn + \frac{D^f_{g0}}{D}\vf $,&
$\displaystyle \vgof = \frac{D^n_{g0\phi}}{D}\vn + \frac{D^f_{g0\phi}}{D}\vf $,
\end{tabular}\hfill(\stepcounter{equation}\theequation,h)
\stepcounter{equation}

\hfill\begin{tabular}{l}
$\displaystyle v_{\rm f, \phi} = \frac{D^n_{e\phi}}{D}\vn + \frac{D^f_{e\phi}}{D}\vf$.
\end{tabular}\hfill(\stepcounter{equation}\theequation)
\end{mathletters}

\noindent
Note that for all $r-$components of the velocities, $D^f_s+D^n_s =D$. 
This can be easily seen if in the determinant $D^f_s+D^n_s$ (which 
is the same as the determinant we get if we substitute the
$s-$column by $\bvec{C^f}+\bvec{C^n}$) we add to the $s-$column the 3
{\em other} columns that also correspond to $r-$velocities. The result
is always the original determinant of the coefficients, $D$.
This is a result of the property of the system of equations in which $r-$velocities, 
whether known or unknown, always appear as velocity
{\em differences} in the equations. This is {\em not} the case for the 
$\phi-$components of the velocities, where the assumption
$v_{n\phi}=0$ breaks the corresponding symmetry. This  
result is the mathematical expression of the physical property of
the $r-$velocities of species, namely, that they vary smoothly between the $r-$velocity of the
field lines and the $r-$velocity of the neutrals. When a species is well attached to the
field lines, its $r-$velocity is equal to that of the
magnetic field lines. When collisions with the neutrals become frequent enough
for the species to fully detach from the field lines, its $r-$velocity
is equal to that of the neutrals.

We can further emphasize this physical picture by defining 
$\Lambda_s = (D^f_s+D^n_s)/D$, $\Theta_s = D^f/D^n$, and noting
that, as we explained above, $\Lambda_s=1$ for all $r-$ velocities. We can
then rewrite the solution of the system as 

\begin{mathletters}
\hfill\begin{tabular}{ll}
$\displaystyle \ve = \frac{1}{\Theta_e+1}\vn + 
\frac{\Theta_e}{\Theta_e+1}\vf\label{soleap}$,&
$\displaystyle \vef =  \Lambda_{e\phi}\left[\frac{1}{\Theta_{e\phi}+1}\vn + 
\frac{\Theta_{e\phi}}{\Theta_{e\phi}+1}\vf\right]\label{solffap}$,
\label{thevelsf}\end{tabular}\hfill(\stepcounter{equation}\theequation,b)
\stepcounter{equation}

\hfill\begin{tabular}{ll}
$\displaystyle \vi = \frac{1}{\Theta_i+1}\vn + 
\frac{\Theta_i}{\Theta_i+1}\vf\label{soliap}$,&
$\displaystyle \vif =  \Lambda_{i\phi}\left[\frac{1}{\Theta_{i\phi}+1}\vn + 
\frac{\Theta_{i\phi}}{\Theta_{i\phi}+1}\vf\right]\label{solifap}$,
\end{tabular}\hfill(\stepcounter{equation}\theequation,d)
\stepcounter{equation}

\hfill\begin{tabular}{ll}
$\displaystyle \vgm = \frac{1}{\Theta_{g-}+1}\vn + 
\frac{\Theta_{g-}}{\Theta_{g-}+1}\vf\label{solmap}$,&
$\displaystyle \vgmf = \Lambda_{g-\phi}\left[\frac{1}{\Theta_{g-\phi}+1}\vn + 
\frac{\Theta_{g-\phi}}{\Theta_{g-\phi}+1}\vf\right]\label{solmfap}$,
\end{tabular}\hfill(\stepcounter{equation}\theequation,f)
\stepcounter{equation}

\hfill\begin{tabular}{ll}
$\displaystyle \vgp=  \frac{1}{\Theta_{g+}+1}\vn + 
\frac{\Theta_{g+}}{\Theta_{g+}+1}\vf\label{solpap}$,&
$\displaystyle \vgpf = \Lambda_{g+\phi}\left[\frac{1}{\Theta_{g+\phi}+1}\vn + 
\frac{\Theta_{g+\phi}}{\Theta_{g+\phi}+1}\vf\right]\label{solpfap}$,
\end{tabular}\hfill(\stepcounter{equation}\theequation,h)
\stepcounter{equation}

\hfill\begin{tabular}{ll}
$\displaystyle \vgo = \frac{1}{\Theta_{g0}+1}\vn + 
\frac{\Theta_{g0}}{\Theta_{g0}+1}\vf\label{solnap}$,&
$\displaystyle \vgof =  \Lambda_{g0\phi}\left[\frac{1}{\Theta_{g0\phi}+1}\vn + 
\frac{\Theta_{g0\phi}}{\Theta_{g0\phi}+1}\vf\right]\label{solnfap}$,
\end{tabular}\hfill(\stepcounter{equation}\theequation,j)
\stepcounter{equation}

\hfill\begin{tabular}{l}
$\displaystyle v_{\rm f \phi} =  \Lambda_{f\phi}\left[\frac{1}{\Theta_{f\phi}+1}\vn + 
\frac{\Theta_{f\phi}}{\Theta_{f\phi}+1}\vf\right]\label{solefap}$.
\end{tabular}\hfill(\stepcounter{equation}\theequation)
\end{mathletters}
\noindent

For the $r-$velocities $\Theta_s$ is then the {\em attachment
  parameter} (i.e., for $\Theta_s \gg 1$, $v_s \approx \vf$ and species $s$ is attached 
to the field lines, while for
$\Theta_s \ll 1$, $v_s \approx \vn$ and species $s$ is detached and comoves with the neutrals).

Finally, $v_f$ can be obtained as a function of $v_n$ by substituting
equations (\ref{thevelsf}a) -(\ref{thevelsf}f) in equation (\ref{ohmf}).

\end{document}